\documentclass[useAMS,usenatbib,referee]{mn2e}
\usepackage{graphicx}
\usepackage{color}

\title[Detection of gamma-ray emission from globular clusters]
  {Detection of gamma-ray emission from globular clusters M 15, NGC 6397, NGC 5904, NGC 6218 and NGC 6139 with Fermi-LAT}
\author[P. F. Zhang et al.]
  {P.~F.~Zhang$^{1,2}$, Y.~L.~Xin$^{2,3}$, L.~Fu$^2$, J.~N.~Zhou$^4$, J.~Z.~Yan$^2$, Q.~Z.~Liu$^2$ and L.~Zhang$^1$.\\
  $^{1}$Department of Physics, Yunnan University, Kunming, China\\
  $^{2}$Key Laboratory of Dark Matter and Space Astronomy, Purple Mountain Observatory, Chinese Academy of Sciences, Nanjing 210008, China\\
  $^{3}$University of Chinese Academy of Sciences, Yuquan Road 19, Beijing, 100049, China \\
  $^{4}$Shanghai Astronomical Observatory, Chinese Academy of Sciences, 80 Nandan Road, Shanghai 200030, China\\}

\begin{document}

\label{firstpage}
\maketitle

\begin{abstract}
In the third Fermi catalogue (3FGL) there are sixteen gamma-ray globular clusters. After analyzing the recent released Pass 8 data of Fermi Large Area Telescope (LAT), we report the discovery of significant gamma-ray emission from M 15 and NGC 6397, confirm that NGC 5904 is a gamma-ray emitter and find evidence of gamma-ray emission from NGC 6218 and NGC 6139. Interestingly, in the globular clusters M 15, NGC 6397 and NGC 5904, millisecond pulsars (MSPs) have been found in radio or X-rays, which are strongly in support of the MSP origin of the gamma-ray emission.  However, due to the relative low luminosity of the gamma-ray emission we do not find any evidence for the gamma-ray pulsation or flux variability of these sources.
\end{abstract}


\section{INTRODUCTION}
Globular clusters (GCs) are nearly the oldest spherical stellar systems with ages greater than $10^{10}$ years in the Milky Way Galaxy and more than 150 GCs \footnote{http://physwww.physics.mcmaster.ca/$\sim$harris/mwgc.dat} have been detected in radio and/or optical bands. The gamma-ray GCs however are much rare. Before the successful performance of Fermi Larger Area Telescope \citep[LAT,][]{Atwood2009}, it was essentially unclear whether GCs consist of a new kind of gamma-ray emitters or not.
GCs are composed of a large number of stars, with higher stellar densities and more compactness in their cores. Through frequent dynamical interactions \citep{verbunt, cheng}, GCs have high stellar encounter rates, and are expected to form low-mass X-ray binaries \citep[LMXBs,][]{clark, katz}. LMXBs are progenitors of millisecond pulsars \citep[MSPs,][]{alpar}. MSPs are believed to emit gamma-ray photons through curvature radiation in the pulsar magnetosphere model or Inverse Compton Scattering (ICS) between surrounding soft photons (such as background star photons in the clusters, the galactic infrared photons, the cosmic microwave background photons and the galactic star photons) and the energetic electrons/positrons \citep[e.g.,][]{cheng2, wei, zhang, harding, venter, bednarek}. It is generally believed that a large group of MSPs contribute to the majority of gamma-ray emission of GCs \citep{abdo2010c}. At the distance of several kpc, the gamma-ray flux of an individual MSP \citep{abdo2009a} is usually inferior to the point-source sensitivity of  Fermi-LAT and it can not be detected. Hence the gamma-rays from GC are most likely contributed by an entire population of MSPs rather than a single one.

The {\it Fermi Gamma-ray Space Telescope}  was launched on June 11, 2008 \citep{Atwood2009} and it scans the whole sky every three hours in survey mode. The LAT onboard Fermi satellite is a pair-production telescope designed to collect gamma-rays in the energy range from about 20 MeV to above 500 GeV with unprecedented sensitivity.
It has been running over seven years and has revolutionized our understanding of the gamma-ray emission from the GCs.  In 3FGL \citep{acero} there are sixteen GCs, including 47 Tucanae \citep{abdo2009a}, Terzan 5 \citep{kong}, Omega Cen\citep{abdo2010a, abdo2010c}, NGC 6388, M 62, NGC 6440, NGC 6652 \citep{abdo2010c}, NGC 6752, M 80, NGC 6541 \citep{tam}, NGC 6624 \citep{tam, freire},  2MS-GC01 \citep{nolan2012, zhou}, NGC 6717, NGC 6441, NGC 6316 and NGC 2808 \citep{acero}. It is estimated that 80\% of the detected MSPs are inside GCs \footnote{http://www.naic.edu/$\sim$pfreire/GCpsr.html}. In June 2015 the Fermi collaboration released their first Pass 8 data. The changes included new pattern recognition algorithm in the Tracker reconstruction, a clustering stage in the Calorimeter reconstruction, better energy reconstruction, more gamma-ray events particularly in the low energy range and so on \footnote{http://fermi.gsfc.nasa.gov/ssc/data/analysis/documentation/
Cicerone/Cicerone\_Data/LAT\_DP.html}. Motivated by such remarkable improvements, in this work we re-analyze the GC sample of \citet{zhou} to search for new gamma-ray emitters.

This work is organized as follows: in Section 2, we describe the data analysis; in Section 3, we present the main results; a summary with some discussions is given in Section 4.

\section{Fermi-LAT DATA ANALYSIS}

\subsection{Data preparation}
The data in our analysis was collected from August 8th, 2008 to August 27th, 2015 with energy range from 100 MeV to 500 GeV. We used the {\it Fermi Science Tools} version {\tt v10r0p5} package\footnote{http://fermi.gsfc.nasa.gov/ssc/data/analysis/software/} provided by the {\it Fermi Science Support Center}. We chose the Pass 8 data (evclass = 128 \& evtype = 3) and only used the events with zenith angles ${\le}$ 90$^{\circ}$ to minimize the contamination from the Earth Limb. We used {\it gtmktime} to get high-quality data in the good time intervals with expression recommended by the LAT team of {\tt (DATA\_QUAL$>$0)\&\&(LAT\_CONFIG==1)}. The instrumental response function (IRF) we adopted here is "P8R2{\_}SOURCE{\_}V6". To reduce the data, we follow the data analysis thread provided by the {\it Fermi Science Support Center}\footnote{http://fermi.gsfc.nasa.gov/ssc/data/analysis/scitools/}.

\subsection{Analysis method}
The events are selected from a 20$^{\circ}\times$20$^{\circ}$ squared region of interest (ROI) with centered coordinates listed Table~\ref{para_source} (just for the sources with significant gamma-ray signal). We binned the data with 25 logarithmical energy bins with spatial pixel size of $0.1^\circ\times0.1^\circ$. Using the binned maximum likelihood method, we modeled the events composed of all known sources included in 3FGL and two diffuse backgrounds in the field. The two background templates include Galactic (gll\_iem\_v06.fits) and extragalactic (iso\_P8R2\_SOURCE\_V6\_v06.txt) diffuse emissions, and we kept normalizations of them as free parameters. The model file was generated with  {\tt make3FGLxml.py}\footnote{http://fermi.gsfc.nasa.gov/ssc/data/analysis/user/}. We used a single power-law (PL, ${dN}/{dE} = N_{0}E^{-\Gamma}$, $N_{0}$ is the normalization and $\Gamma$ is the photon spectral index) or an exponential cut-off power-law (PLE, ${dN}/{dE} = N_{0}E^{-\Gamma}e^{-{E}/{E_{c}}}$, $E_{c}$ is the cut-off energy) as the spectral model of gamma-ray GC candidates. Within 5$^\circ$ from the center of ROI, the normalization and the spectral parameters of each source were set free. While for other sources located in the ROI, we only free the normalization factors.

\begin{figure*}
\centering
	\includegraphics[scale=0.35]{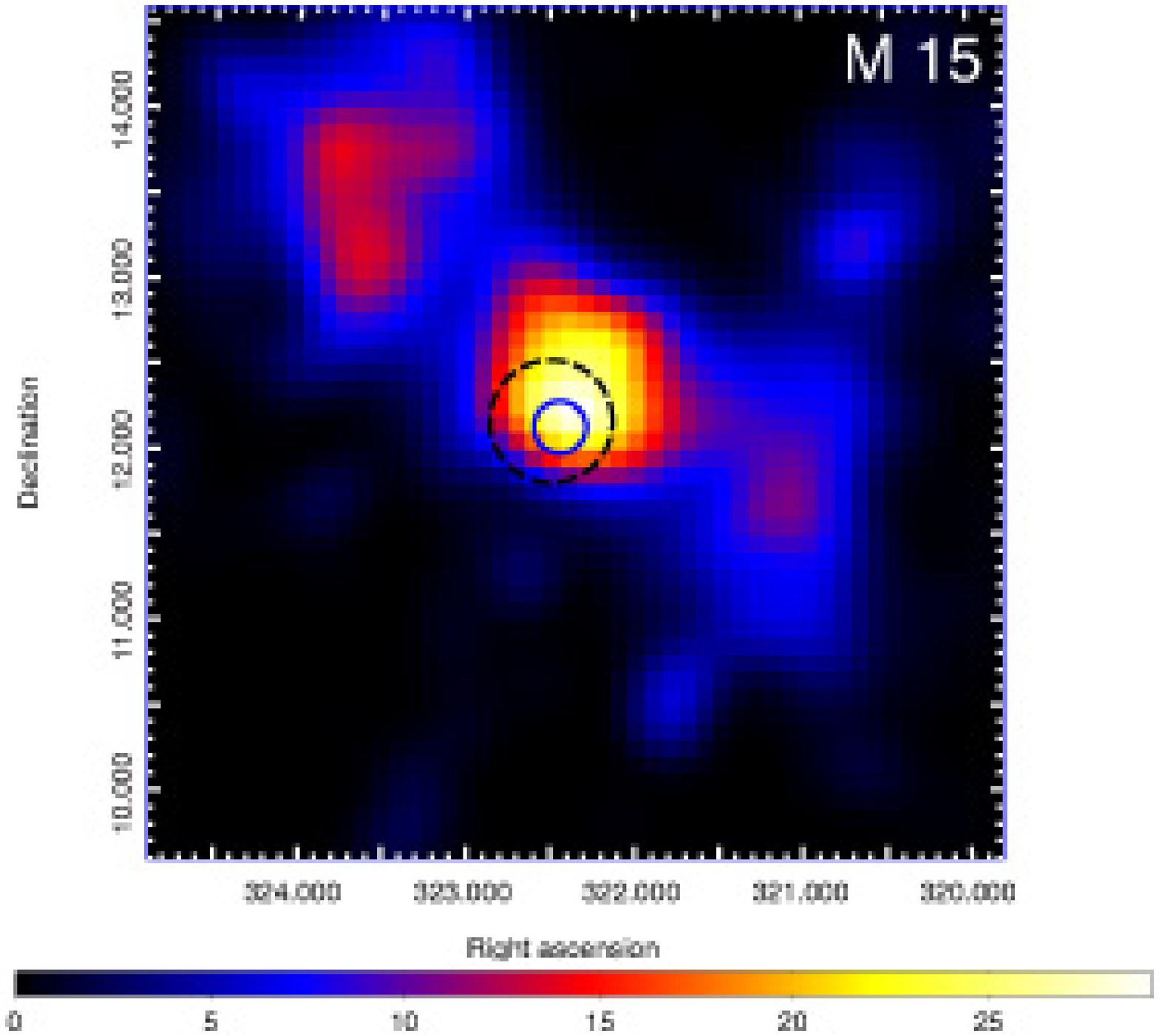}
	\includegraphics[scale=0.35]{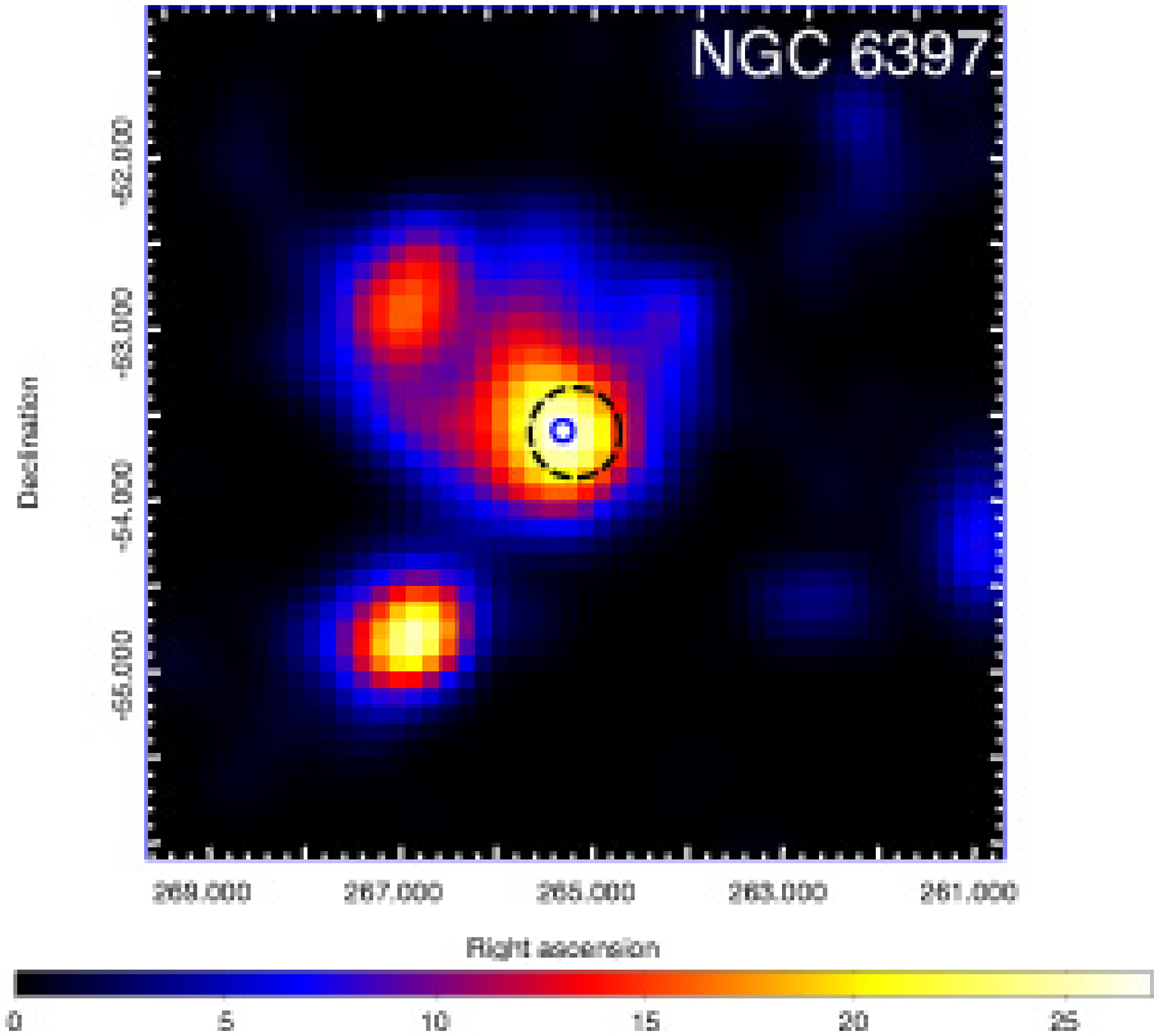}
	\includegraphics[scale=0.35]{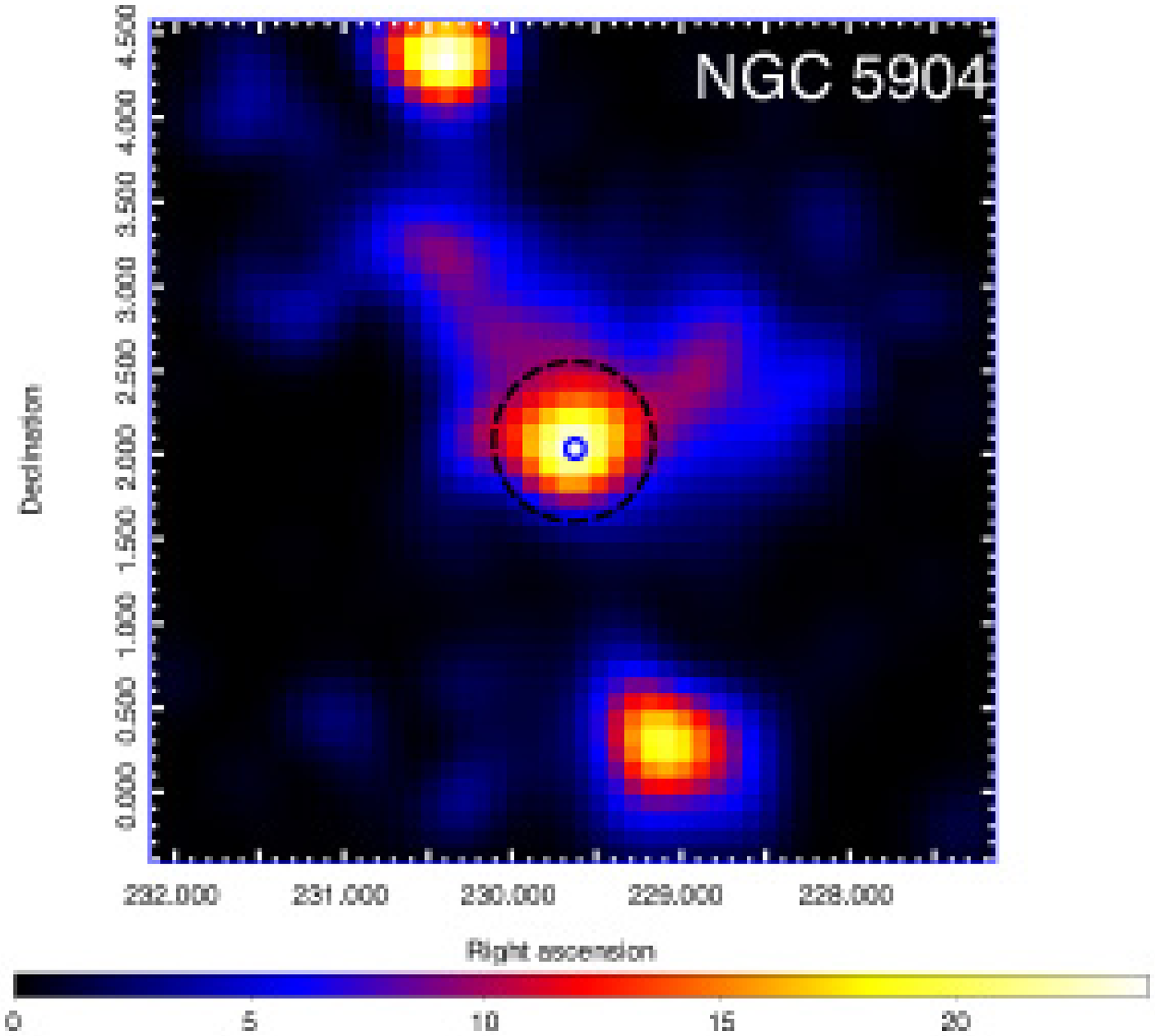}
	\includegraphics[scale=0.35]{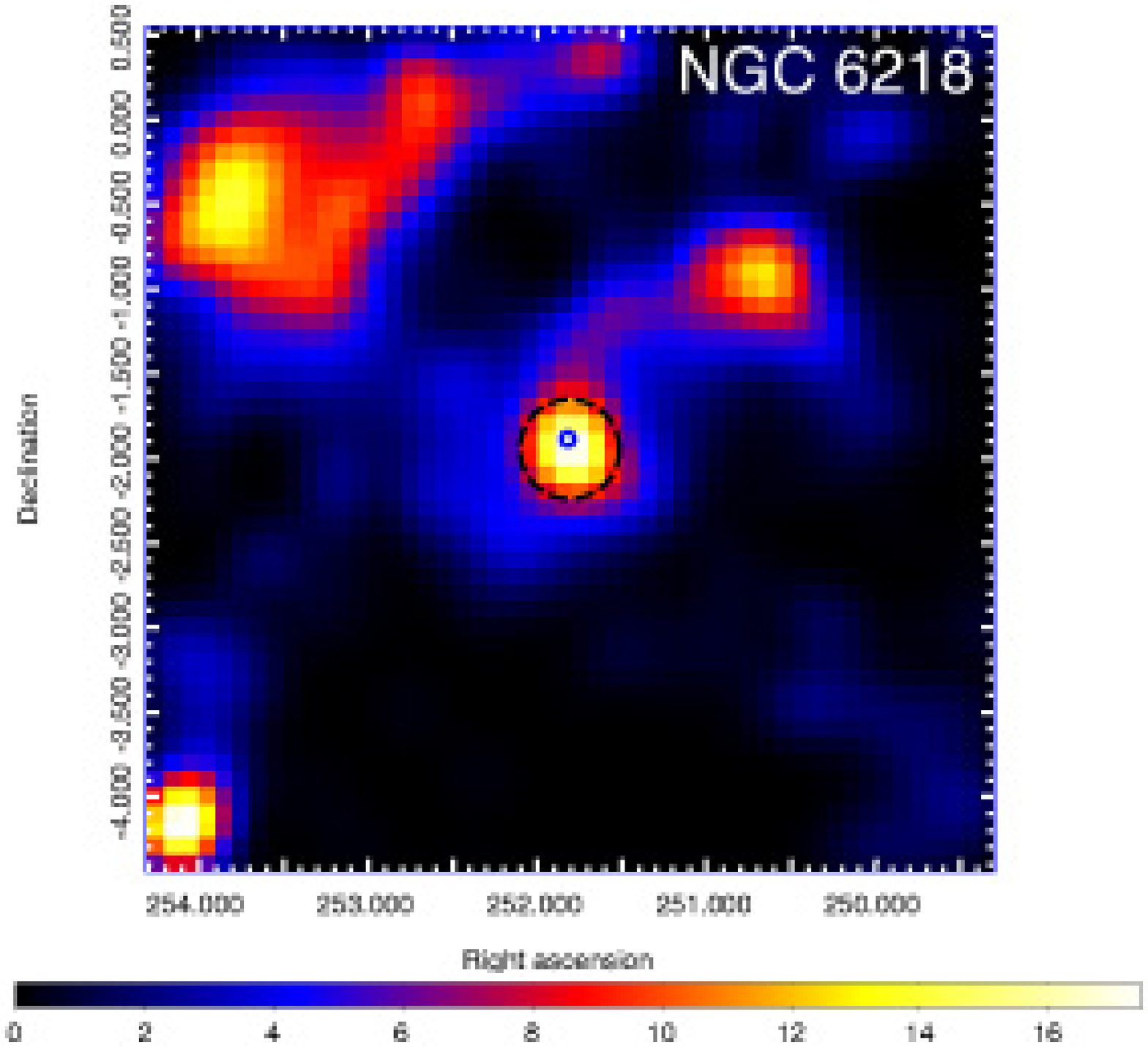}
	\includegraphics[scale=0.35]{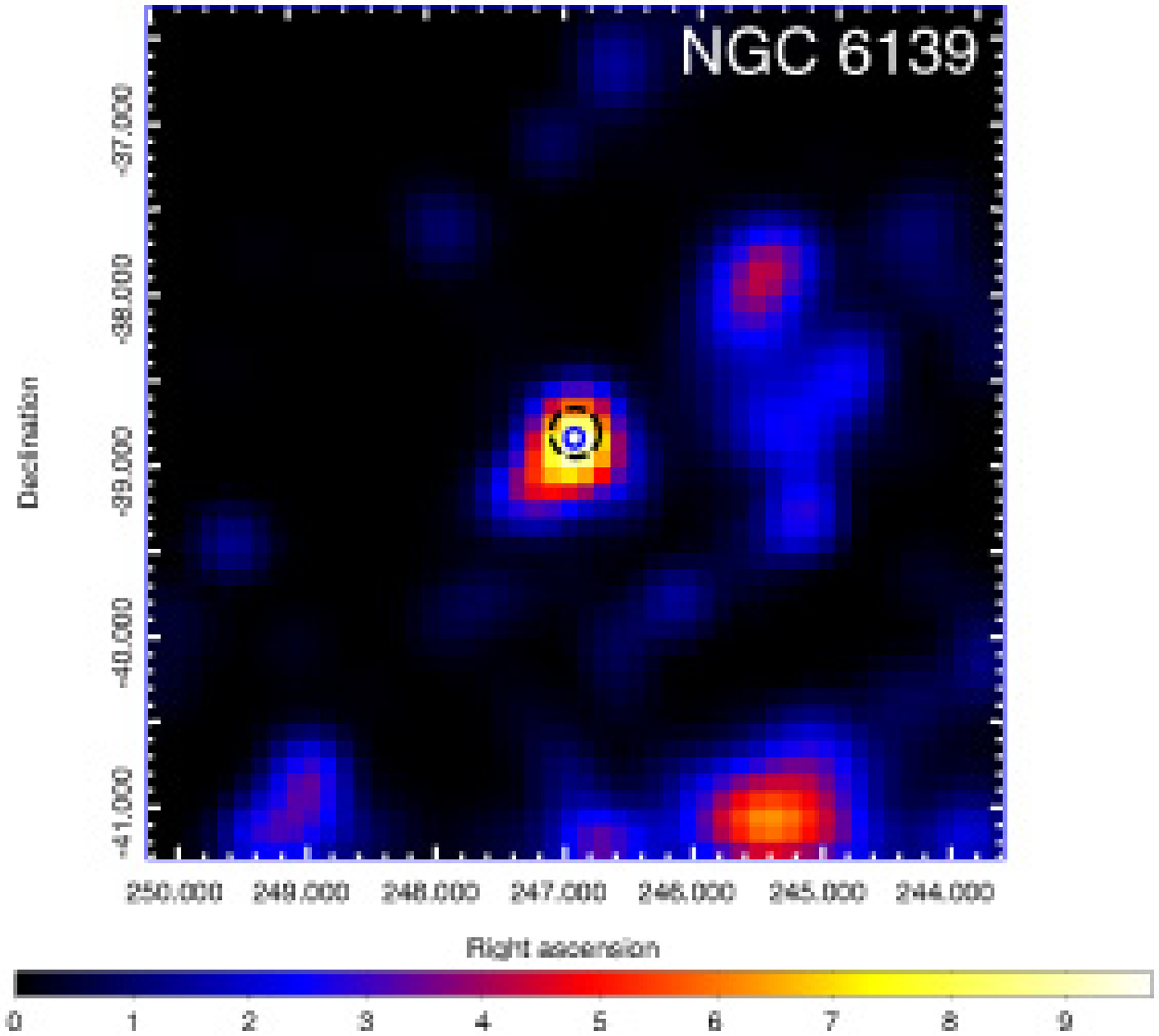}
		\caption{TS maps of 5$^{\circ}\times$5$^{\circ}$ region for M 15, NGC 5904, NGC 6218, NGC 6139 and NGC 6397. The blue solid circles represent the best-fit centroid of the gamma-ray emission and their radii indicate the 1$\sigma$ statistical error. The black dash circles represent the tidal radii \citep[][ 2010 version]{harris} centered on the nominal locations of five globular clusters. These maps are smoothed by Gaussian function whose kernel radius is 0.3$^\circ$.}
	\label{TSmaps}
\end{figure*}

\citet{zhou} carried out an systematic search of gamma-ray GCs. In addition to the discovery of a few gamma-ray emitters, they found 31 sources with weak signals. In this work, we re-analyze these tentative sources and focus on five sources with
test statistic (TS) values greater than 25 (i.e., the corresponding significance is $\sim4\sigma$, Abdo et al. 2010b) and normal photon spectral indexes (i.e., $\Gamma\leq 3$). The TS value is defined as TS$=-2(\mathcal{L}_0 - \mathcal{L}_1)$, where $\mathcal{L}_0$ and $\mathcal{L}_1$ are the logarithmic maximum likelihood values of null hypothesis and tested model including the target source. We only find five candidates, including NGC 5904, M 15, NGC 6218, NGC 6139 and NGC 6397, for further analysis. We then use {\it gttsmap} to generate their $5^\circ\times5^\circ$ residual TS maps (see Figure~\ref{TSmaps}, each map is centered on the nominal position of the GC from SIMBAD\footnote{http://simbad.u-strasbg.fr/simbad/}). The {\it gttsmap} removes all sources that within model file through a grid of locations on the sky and maximizing $-log(Likelihood)$ at each grid positions to identify weaker sources. We select events above 400 MeV for the residual TS maps. In running {\it gttsmap}, we freeze all parameters of sources at the best-fit results of {\it gtlike} except the two diffuse backgrounds. In \citet{nolan2012}, the source with a $\rm TS_{curve} > 16$ had been considered significantly curved, where $\rm TS_{curve}$ is defined as $\rm TS_{curve} = 2(log~\mathcal{L}(curved~spectrum) - log~\mathcal{L}(power-law))$. But for the current five GCs we have $\rm TS_{curve}\ll 16$. Hence we use a power-law spectrum as the best-fit model for each GC (Note that we have also presented the PLE result for NGC 5904 to compare with the result of \citet{zhou}).

Using {\it gtfindsrc} we derive the positions of the gamma-ray sources, which are then found to be within the tidal radii of NGC 5904, M 15, NGC 6218, NGC 6139 and NGC 6397, respectively (see Fig.\ref{TSmaps}). The association of these five gamma-ray emitters, not included in 3FGL, with the GCs is established.
Following \citet{tam}, we estimated the trial factor to avoid fake signal with a number of search positions. In these 31 GC candidates, we try to search the gamma-ray emissions and only five of them are further analyzed in this work. So $N_{GC}$ is 31, and we use an averaged tidal radius $13'$; $N_{bin}$ is about 15 which is derived by dividing the trial factor area of GC into 0.1$^{\circ}\times$0.1$^{\circ}$ pixel. Our best-fit results are summarized in Table \ref{para_spec} with statistical errors only. Finally we can get the post-trail significance (see Table \ref{para_spec}).

To obtain the spectral energy distributions (SEDs) with $gtlike$, we divided the energy range into equal logarithmic spaced energy bins of these five GCs. In this step, we forze all spectral parameters. By fitting all model components, we got the flux of target source in each energy bin. We have also examined possible flux variability, in which all data were binned into 10 equal time bins. For each time bin, the flux of target source was derived using $gtlike$. Using a $\chi^2$ minimization procedure in this analysis, we fit the lightcurves with a constant flux model. Since the dataset are binned into 10 bins, we have 9 degrees of freedom for each globular cluster. If the value of reduced $\chi^2_{min}$ above 25.26, we will claim that a significant variability of flux was detected at the 99.73\% confidence levels (i.e., 3$\sigma$). The results of this minimization are revealed in Section 3.

\section{Results}
\subsection{Significant $\gamma-$ray detections}
\subsubsection{NGC 5904}
\begin{figure*}
\centering
	\includegraphics[scale=0.45]{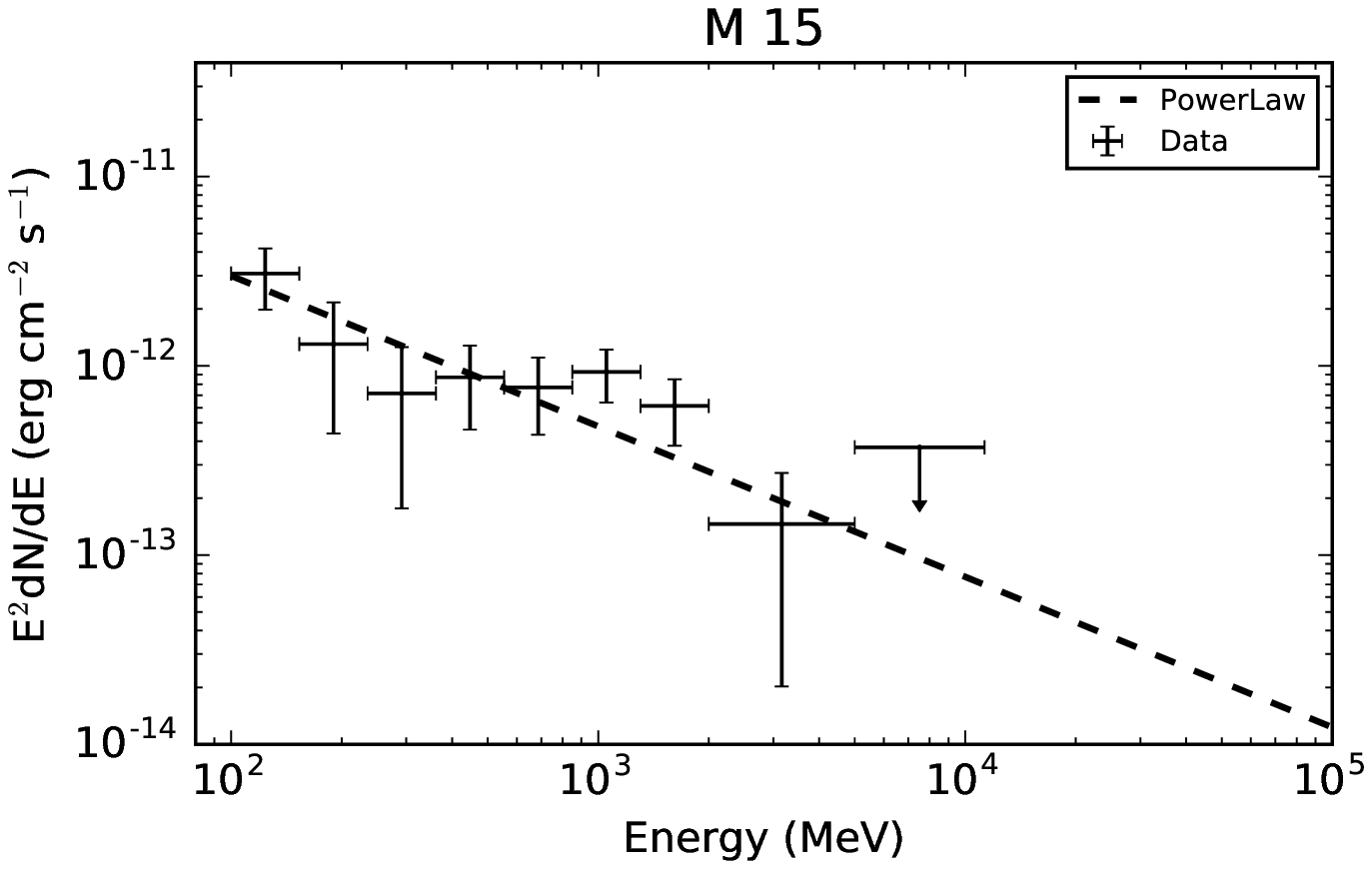}
	\includegraphics[scale=0.45]{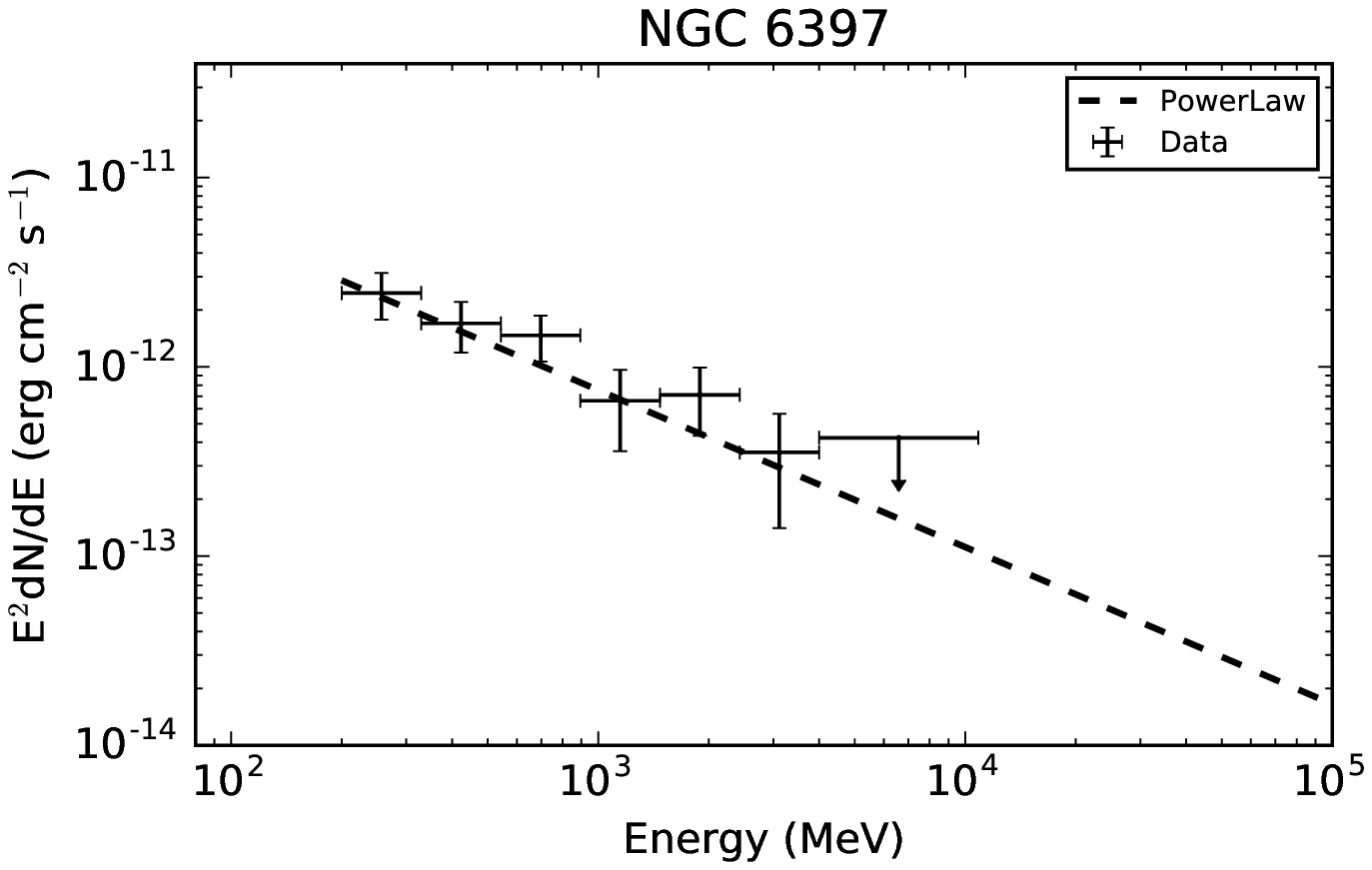}
	\includegraphics[scale=0.45]{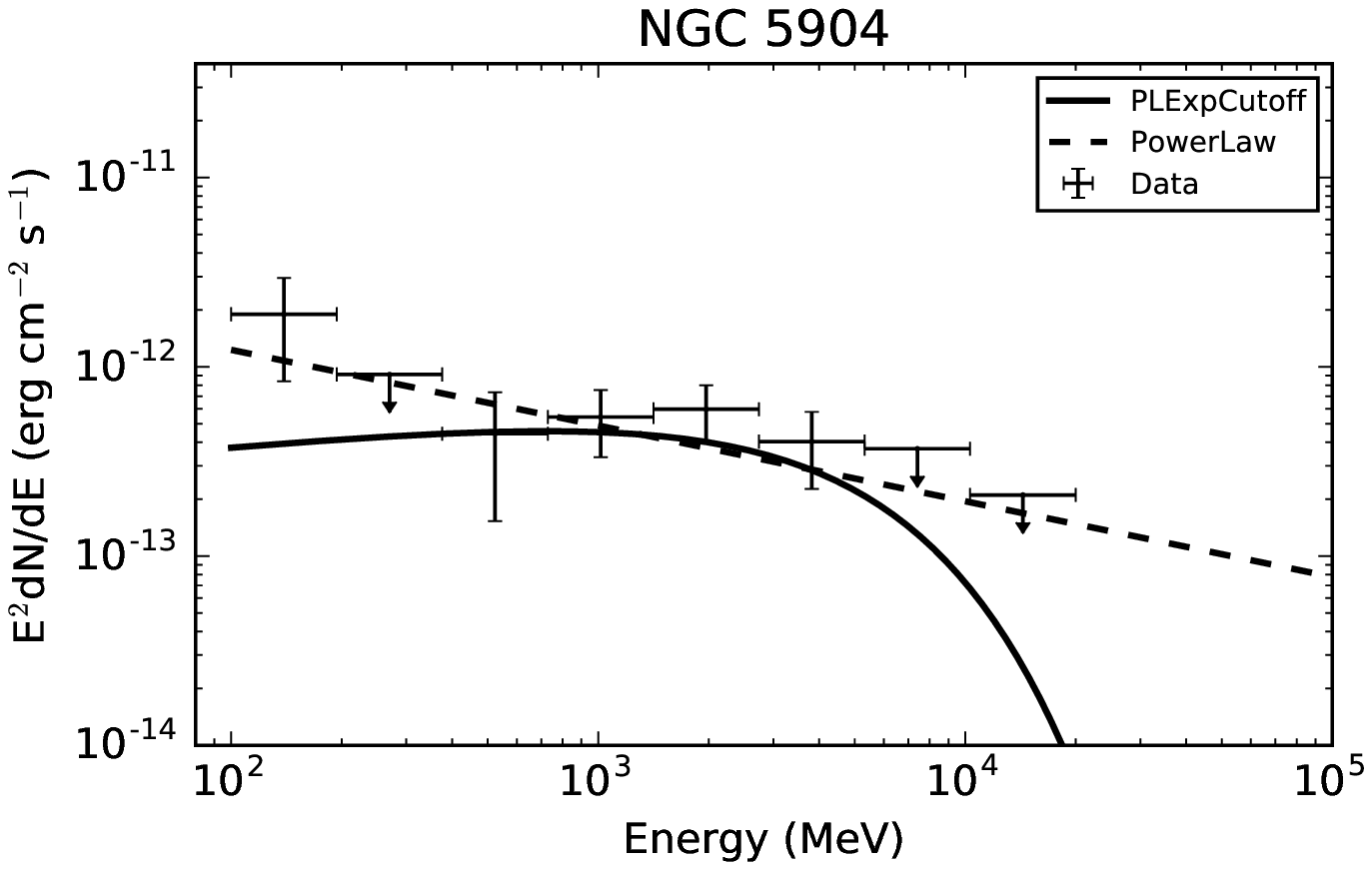}
	\includegraphics[scale=0.45]{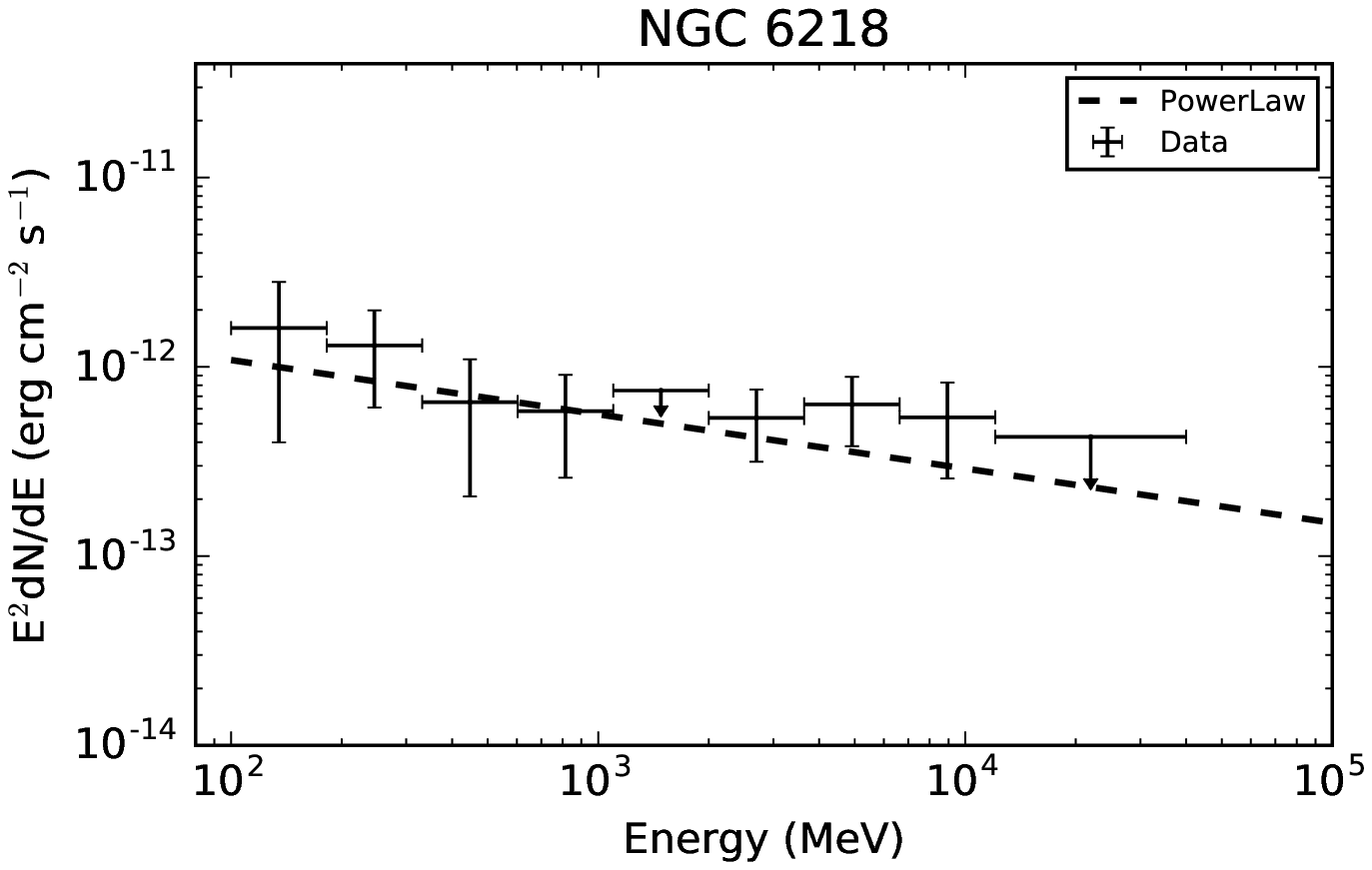}
	\includegraphics[scale=0.45]{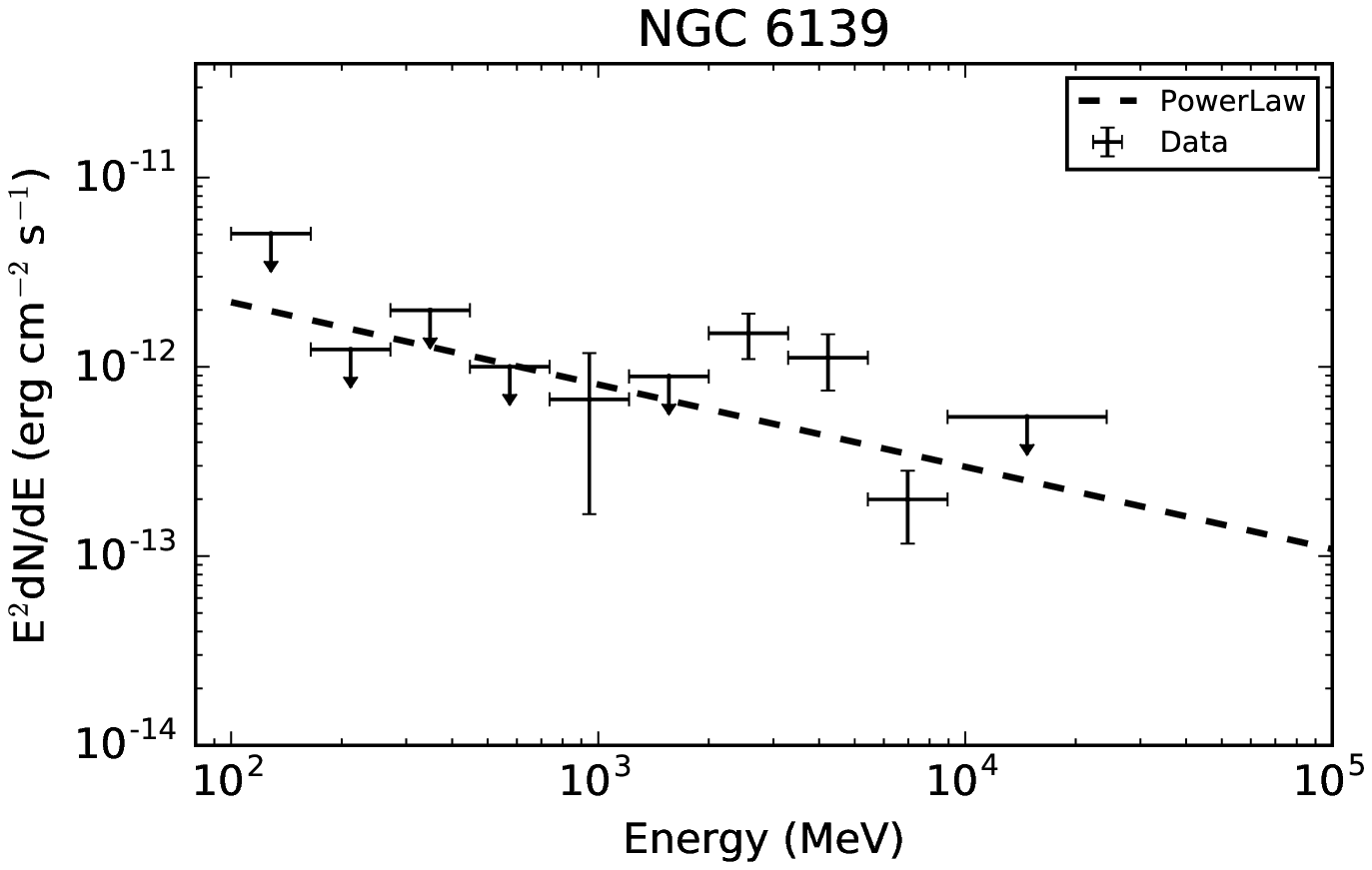}
		\caption{SEDs of the five globular clusters. The dash lines indicate the best fitted spectral model ($E^2\times dN/ dE$), while the solid lines indicate spectral fitting with a plausible model. The spectral parameters of the models are quoted in Table~\ref{para_source}.}
	\label{SED}
\end{figure*}

Its tidal radius is $12.07'$. The distance from the Sun is 7.5 kpc. Five pulsars were found in this GC \citep{anderson, hessels}. With the Pass 8 data, we confirmed \citet{zhou} that NGC 5904 is a gamma-ray emitter.
We fit the events above 100 MeV of this GC with a power law spectrum model and get the best-fitting photon spectral index $\Gamma$ of 2.40 $\pm$ 0.14. The TS value of NGC 5904 is 44, corresponding to a detection significance of 6.6$\sigma$ (i.e., 5.3$\sigma$ post-trial). The best-fitting position (the results of $gtfindsrc$) of NGC 5904 is RA = 229.63$^\circ$ and Dec = 2.04$^\circ(\pm~0.06^\circ$, 1$\sigma$ error) (J2000), which is offset by $2.65$ arcmin from the nominal position (RA = 229.64$^\circ$, Dec = 2.08$^\circ$), well within its tidal radius. The integrated photon flux above 100 MeV is $F_{\rm 0.1-500~GeV} = (5.94 \pm 1.87)\times10^{-9}~{\rm cm}^{-2} ~{\rm s}^{-1}$ and the integrated energy flux is $E_{\rm 0.1-500~GeV} = (3.22 \pm 0.60) \times 10^{-12}~{\rm erg}~{\rm cm}^{-2}~{\rm s}^{-1}$; the gamma-ray luminosity is $L_{\rm 0.1-500~GeV} = (2.18 \pm 0.40) \times 10^{34}~{\rm erg~ s^{-1}}$. These best-fit results are consistent with that of \citet{zhou}. The reduced $\chi^2_{min}$ of minimization of its lightcurve is 16.97/9. We do not find evidence for flux variability at the confidence level of 3$\sigma$.
The residual TS map above 400 MeV of this GC is shown in Figure~\ref{TSmaps} (Upper left panel) and its SED is shown in Figure~\ref{SED} (Upper left panel).

\subsubsection{M 15}
There are eight radio pulsars in this GC. The tidal radius are $21.50'$. Fitting the spectrum with the power law model, we obtain an index ($\Gamma$) of 2.84 $\pm$ 0.18 and the TS value is 49 corresponding to a detection significance of 7.0 $\sigma$ (i.e., 5.7$\sigma$ post-trial). The best-fitting position of M 15 is RA = 322.44$^\circ$, Dec = 12.13$^\circ(\pm0.15^\circ$) (J2000, the nominal position is RA = 322.35$^\circ$, Dec = 12.1$^\circ$), which is offset by 3.72 arcmin from M 15 core, well within the tidal radius of M 15. The integrated photon flux above 100 MeV is $F_{\rm 0.1-500~GeV} = (11.84\pm2.48)\times10^{-9}~{\rm cm}^{-2} ~{\rm s}^{-1}$ and the integrated energy flux is $E_{\rm 0.1-500~GeV} = (4.15 \pm 0.65) \times 10^{-12}~{\rm erg}~{\rm cm}^{-2}~{\rm s}^{-1}$. At a distance of 10.3 kpc, the gamma-ray luminosity is $L_{\rm 0.1-500~GeV} = (5.26^{+~1.31}_{-~1.16}) \times 10^{34}~{\rm erg~s^{-1}}$. With the first year Fermi-LAT data, M 15 was not detected in gamma-ray band by  \citet[][a small TS$=5.4$ was reported]{abdo2010c} and an upper limit was set by adopting a ``typical" PLE spectrum. Thanks to the improved understanding of the Galactic diffuse emission as well as the increase of the observation data, in this work M 15 was successfully identified as a $\gamma$-ray emitter. The residual TS map (between 400 MeV and 500 GeV) of this GC is shown in Figure~\ref{TSmaps} (Upper right panel) and its SED is shown in Figure~\ref{SED} (Upper right panel). Although the best-fit centroid does not exactly trace the pixel with highest TS value due to statistic and systematic noise, the offset is within $1\sigma$ error circle and thus ignorable. The reduced $\chi^2_{min}$ of minimization is 7.70/9, implying the absence of variability at 3$\sigma$ confidence level.

\subsubsection{NGC 6397}
The global cluster  NGC 6397 locates at 2.2 kpc from the Sun in the constellation Ara as one of the two nearest globular clusters (the other is NGC 6121) to us. The age of this GC is estimated to be 13.9 $\pm$ 1.1 billion years old \citep{gratton}, almost the same as the Universe. In this globular cluster, one radio pulsar has been detected. The nominal position is RA = 265.18$^\circ$, Dec = $-53.67^\circ$ ($l$ = $338.17^\circ$, $b$ = $-11.96^\circ$) and our best-fitting position is RA = 265.30$^\circ$, Dec = $-53.66^\circ(\pm~0.06^\circ$), which is offset by 4.4 arcmin from its core, well within the tidal radius of NGC 6397 (the tidal radius is $15.81'$). The power-law spectrum fit yields an index $\Gamma$ of 2.83 $\pm$ 0.17. The TS value of the gamma-ray signal is 49, corresponding to a detection significance of 7.0 $\sigma$ (i.e., 5.7$\sigma$ post-trial). The integrated photon flux between 200 MeV and 500 GeV is $F_{\rm 0.2-500~GeV} = (4.94 \pm 0.88)\times10^{-9}~{\rm cm}^{-2} ~{\rm s}^{-1}$. The integrated energy flux is $E_{\rm 0.2-500~GeV} = (3.47 \pm 0.54) \times 10^{-12}~{\rm erg}~{\rm cm}^{-2}~{\rm s}^{-1}$ and the gamma-ray luminosity is $L_{\rm 0.2-500~GeV} = (0.20^{+~0.15}_{-~0.12}) \times 10^{34}~{\rm erg~s^{-1}}$. The reduced $\chi^2_{min}$ is 11.97/9, implying the absence of variability at 3$\sigma$ confidence level.
The residual TS map above 400 MeV is shown in Figure~\ref{TSmaps} (Middle left panel) and its SED above 200 MeV is shown in Figure~\ref{SED} (Middle right panel).

\subsection{Possible $\gamma-$ray detections}

\subsubsection{NGC 6218}
Its nominal position is RA = 251.81$^\circ$, Dec = $-1.95^\circ$ ($l$ = $15.72^\circ$, $b$ = $26.31^\circ$) and there is no pulsar has been detected in this GC so far. Its tidal radius is $17.6'$. Using {\it gtlike}, we fit the events with a single power law model and obtain a photon index of $\Gamma$ = 2.28 $\pm$ 0.2. Its TS value is 36, corresponding to a detection significance of 6.0$\sigma$ (i.g., 4.5$\sigma$ post-trial). The best-fitting position, RA = 251.82$^\circ$ and Dec = $-1.89^\circ(\pm0.04^\circ$), which is offset by 3.68 arcmin from its core, is well within the tidal radius circle of NGC 6218. Its integrated photon and energy fluxes (from 100 MeV to 500 GeV) are (5.28 $\pm$ 2.44) $\times$ 10$^{-9}$ cm$^{-2}$ s$^{-1}$ and (3.46 $\pm$ 0.73) $\times$ 10$^{-12}$ erg cm$^{-2}$ s$^{-1}$, respectively. The gamma-ray luminosity of NGC 6218 is $L_{\rm 0.1-500 GeV} = (9.54 \pm 2.0) \times 10^{33}~{\rm erg}~{\rm s}^{-1}$ at a distance of 4.8 kpc.
The residual TS map above 400 MeV is shown in Figure~\ref{TSmaps} (Middle right panel) and its SED above 200 MeV is shown in Figure~\ref{SED} (Middle right panel). The reduced $\chi^2_{min}$ of minimization is 9.79/9, showing no evidence for variability at 3$\sigma$ confidence level.

\subsubsection{NGC 6139}
Its tidal radius is $8.52'$. No radio pulsar has been detected in this globular cluster. We find the evidence of gamma-ray emission in the tidal radius of NGC 6139 with TS value of 27 corresponding to a detection significance of 5.2$\sigma$ (i.g., 3.4$\sigma$ post-trial). We fit the events with a single power law model and obtain the photon index $\Gamma$ of 2.44 $\pm$ 0.12. The best-fitting position, RA = 246.92$^\circ$ and Dec = $-38.88^\circ(\pm0.05^\circ$), is offset by 1.73 arcmin from its core, well within the tidal radius circle of NGC 6139. The integrated photon flux between 100 MeV and 500 GeV is $F_{\rm 0.1-500~GeV} = (9.56 \pm 2.42)\times10^{-9}~{\rm cm}^{-2} ~{\rm s}^{-1}$. The integrated energy flux is $E_{\rm 0.1-500~GeV} = (4.93 \pm 0.86) \times 10^{-12}~{\rm erg}~{\rm cm}^{-2}~{\rm s}^{-1}$ and the gamma-ray luminosity is $L_{\rm 0.1-500~GeV} = (5.21^{+~2.28}_{-~1.78}) \times 10^{34}~{\rm erg~ s^{-1}}$ at a distance of 9.4 kpc.
The decrease of the TS value of NGC 6139 in comparison to that reported in \citet{tam} could be due to the ``underestimate" of the Galactic diffuse emission in such a direction in the early version of the Fermi-LAT science tool. Nevertheless, the spectral index as well as the fluence reported in \citet{tam} are similar to ours.
The residual TS map above 400 MeV is shown in Figure~\ref{TSmaps} (Lower panel) and the SED above 100 MeV is shown in Figure~\ref{SED} (Lower panel). The reduced $\chi^2_{min}$ of minimization of lightcurve is 4.98/9, without displaying any evidence for variability at 3$\sigma$ confidence level.

\section{DISCUSSION}
We have systematically analysed the data from 31 GCs and finally found significant gamma-ray emission from five ones. We have discovered gamma-rays from NGC 5904, M 15 and NGC 6397 that have post-trail significance above $5\sigma$. We have also found strong evidence of gamma-ray emission from NGC 6218 and NGC 6139 (for which the post-trial significance are below 5$\sigma$ and more data are needed to draw the final conclusion). Among these five GCs, the most distant object is M 15 (10.3 kpc), while NGC 6397 is the closest one (2.2 kpc). As shown in Fig.4 and Table \ref{para_spec}, we do not find significant differences for the current five gamma-ray GCs from the previous events in terms of gamma-ray luminosity, spectra, number of known MSPs and other cluster properties. Except NGC 6139, the significance of the gamma-ray emission of NGC 5904 and M 15 increased significantly thanks to the longer exposure time and the remarkable improvements of Pass 8 data in comparison to Pass 7 data. The higher TS value of gamma-ray emission from NGC 6139 found in 2011 is likely due to the underestimated diffuse emission background contamination in that direction in previous Fermi Pass 7 data. Since the majority of gamma-ray emission from GCs is widely believed to be from a population of MSPs, we can estimate the number of MSPs in these GCs under the assumption that each MSP emits similar amount of gamma-rays \citep{abdo2009b}.
Following \citet{abdo2009b}, the total number of MSPs in a given GC $N_{\rm MSP}$ is evaluated by $N_{\rm MSP}={L_{\gamma}}/{\langle\dot E \rangle \langle \eta_{\gamma} \rangle}$, where $L_{\gamma}$ is the isotropic gamma-ray luminosity of the cluster; $\langle\dot E \rangle$ is the average spin-down power of MSP and $\langle \eta_{\gamma} \rangle$ is the estimated average spin-down to gamma-ray luminosity conversion efficiency. Following \citet{abdo2009b}, we adopt a value of $\langle\dot E \rangle = (1.8\pm0.7)\times10^{34}$ erg s$^{-1}$ for all GCs and set $\langle \eta_{\gamma} \rangle = 0.08$. With $L_\gamma = 4\pi S d^2$ we get the isotropic gamma-ray luminosity of the cluster, where $S$ is the observed energy flux and $d$ is the distance from the Sun to the GC. These parameters are both summarized in Table~\ref{para_spec}. The results of the estimated number of MSPs in each GC are also shown in Table~\ref{para_spec}, which are more than that identified in radio and optical bands, implying that much more MSPs would be identified in long-wavelength bands in the future.

Using {\it gtlike}, we also separate each data set into 10 equal time bins for searching the possible flux variability for these five globular clusters. No significant variability is obtained. This can be understood since MSPs are stable in timescales much longer than their rotation periods. \citet{freire} reported a high significant gamma-ray pulsation from PSR J1823-3021A in NGC 6624. We also searched for the gamma-ray pulsation in NGC 5904, M 15 and NGC 6397 with the radio pulsar ephemeris. Using Tool TEMPO2 \citep{hobbs, edwards} which is not associated with the FSSC, we fold the events from the three GCs on every known radio pulsar ephemeris. The events are assigned phases base on its arrival time to investigate a pulse profile. We do not find any evidence for the gamma-ray pulsation. This is natural since usually individual MSP is too dim to be detected in gamma-ray band and the detected signal in each GC is most likely from a group of MSPs, for which the pulsation from individual MSP is smeared.

\begin{figure*}
\centering
	\includegraphics[scale=0.4]{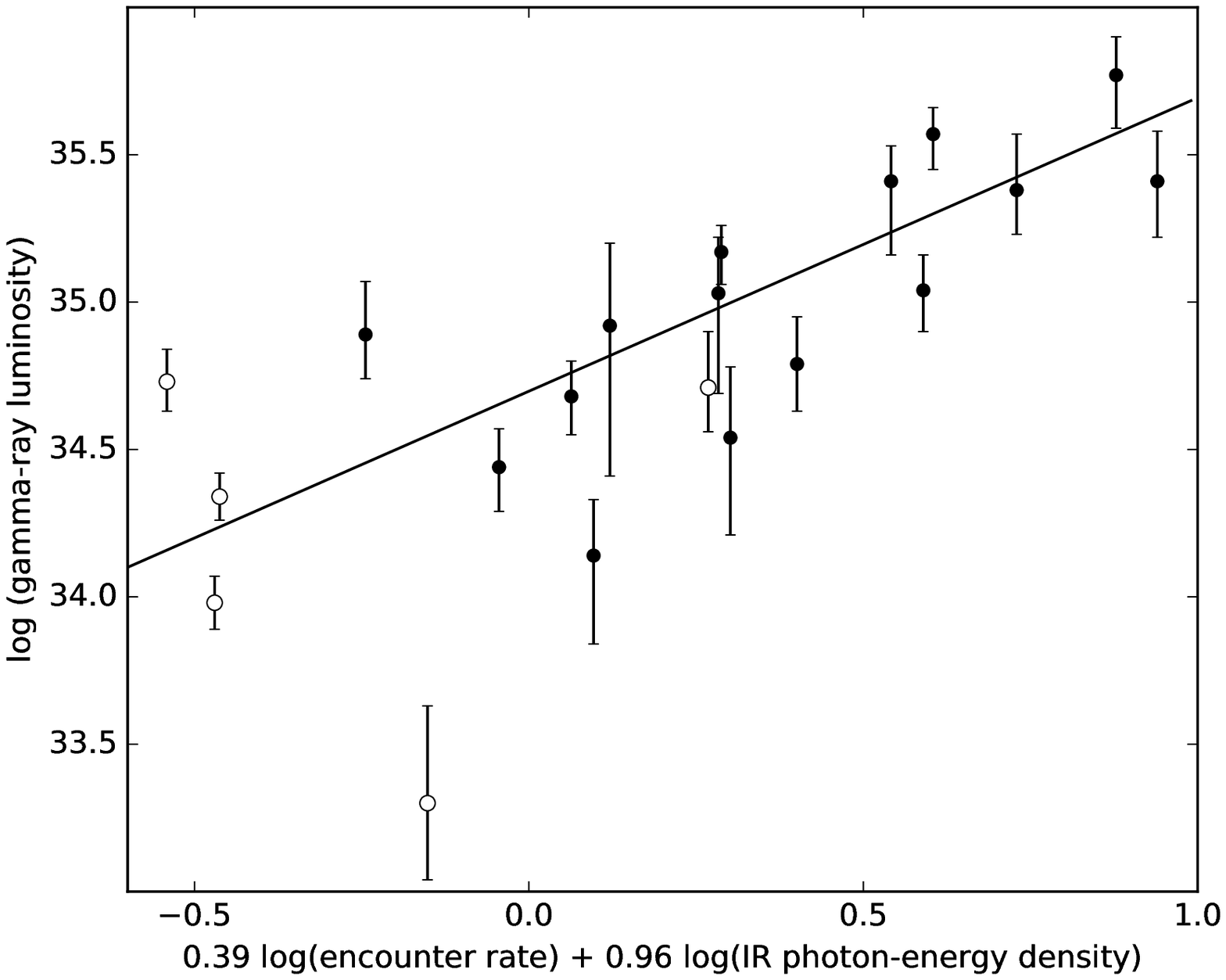}
	\includegraphics[scale=0.4]{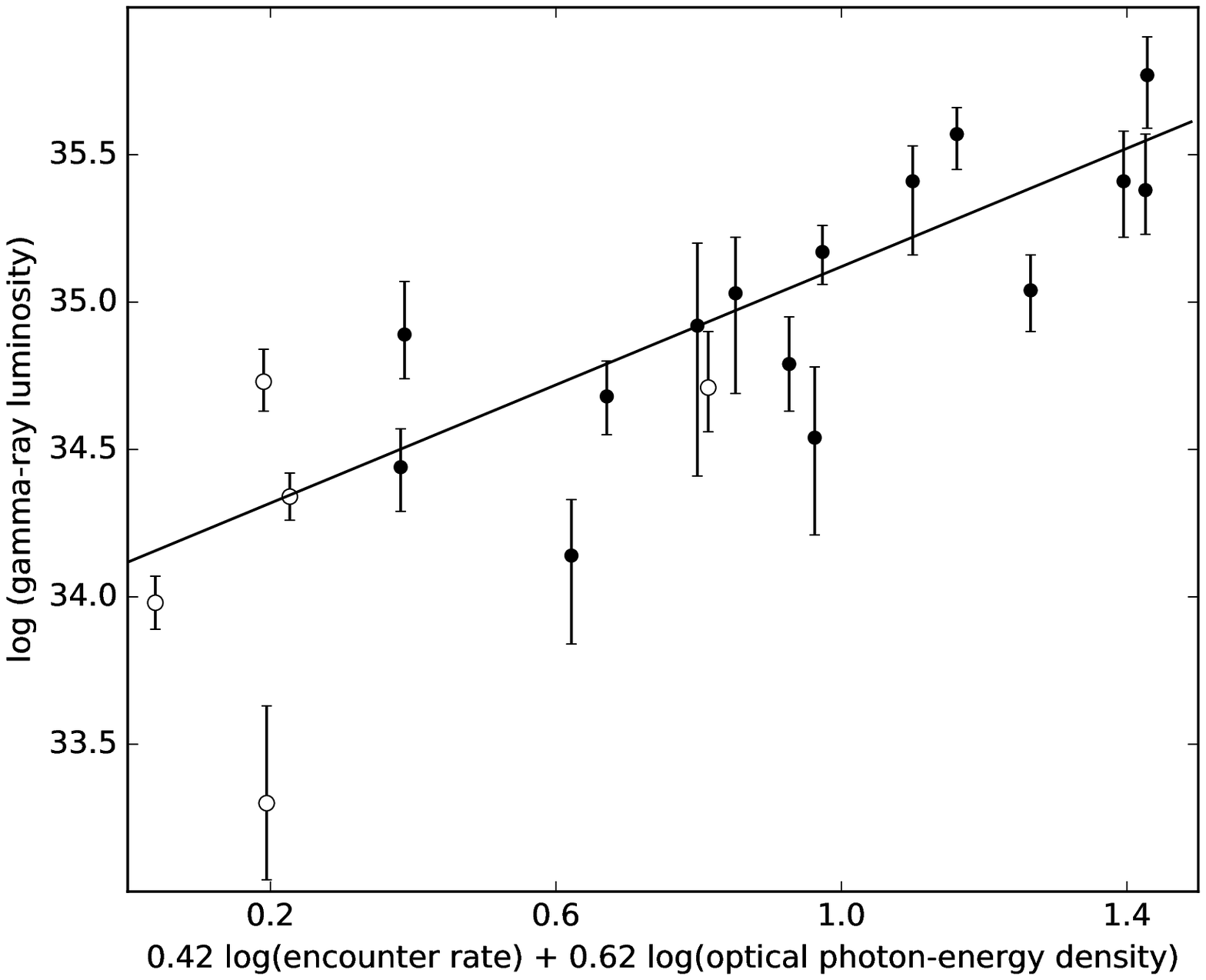}
	\includegraphics[scale=0.4]{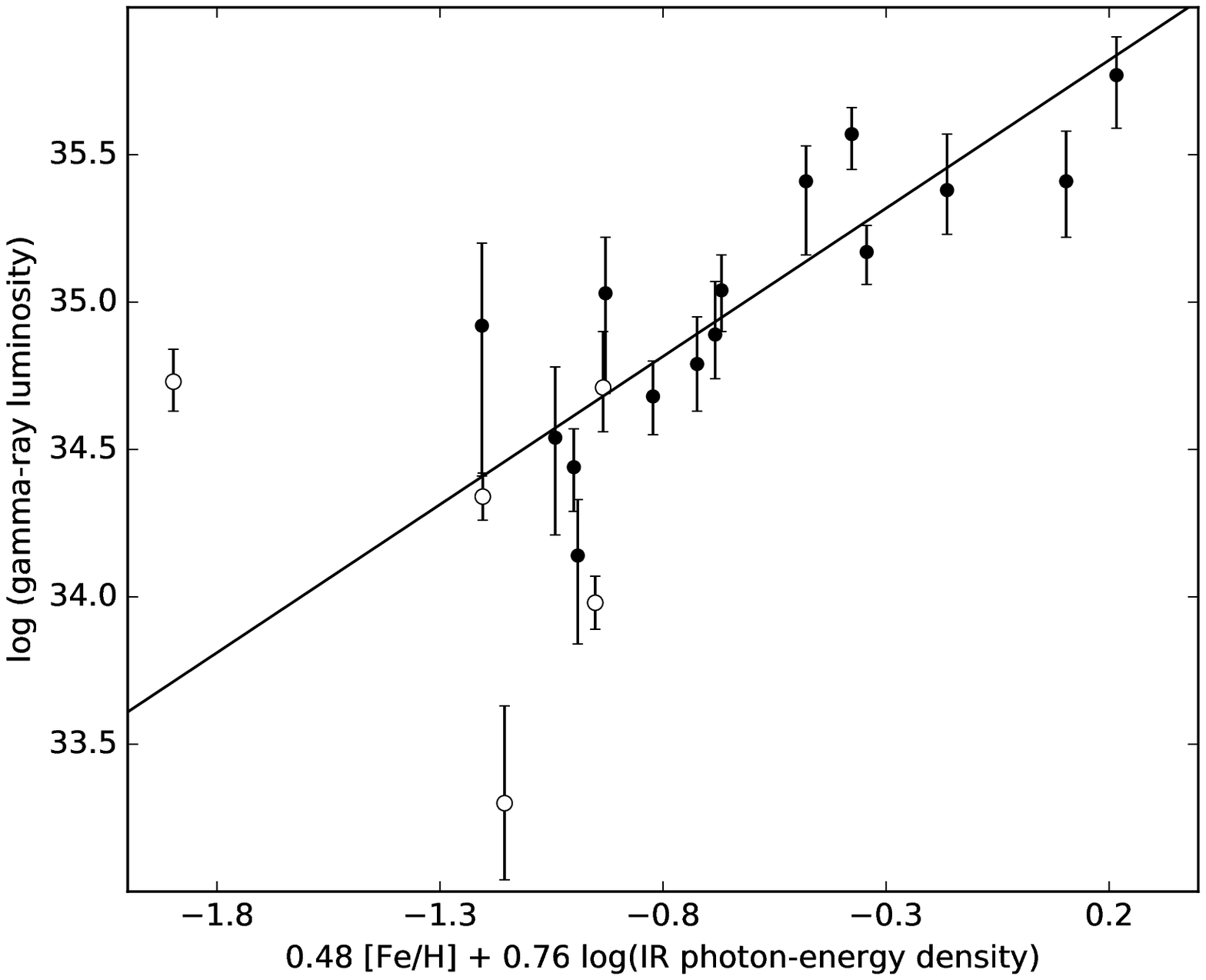}
	\includegraphics[scale=0.4]{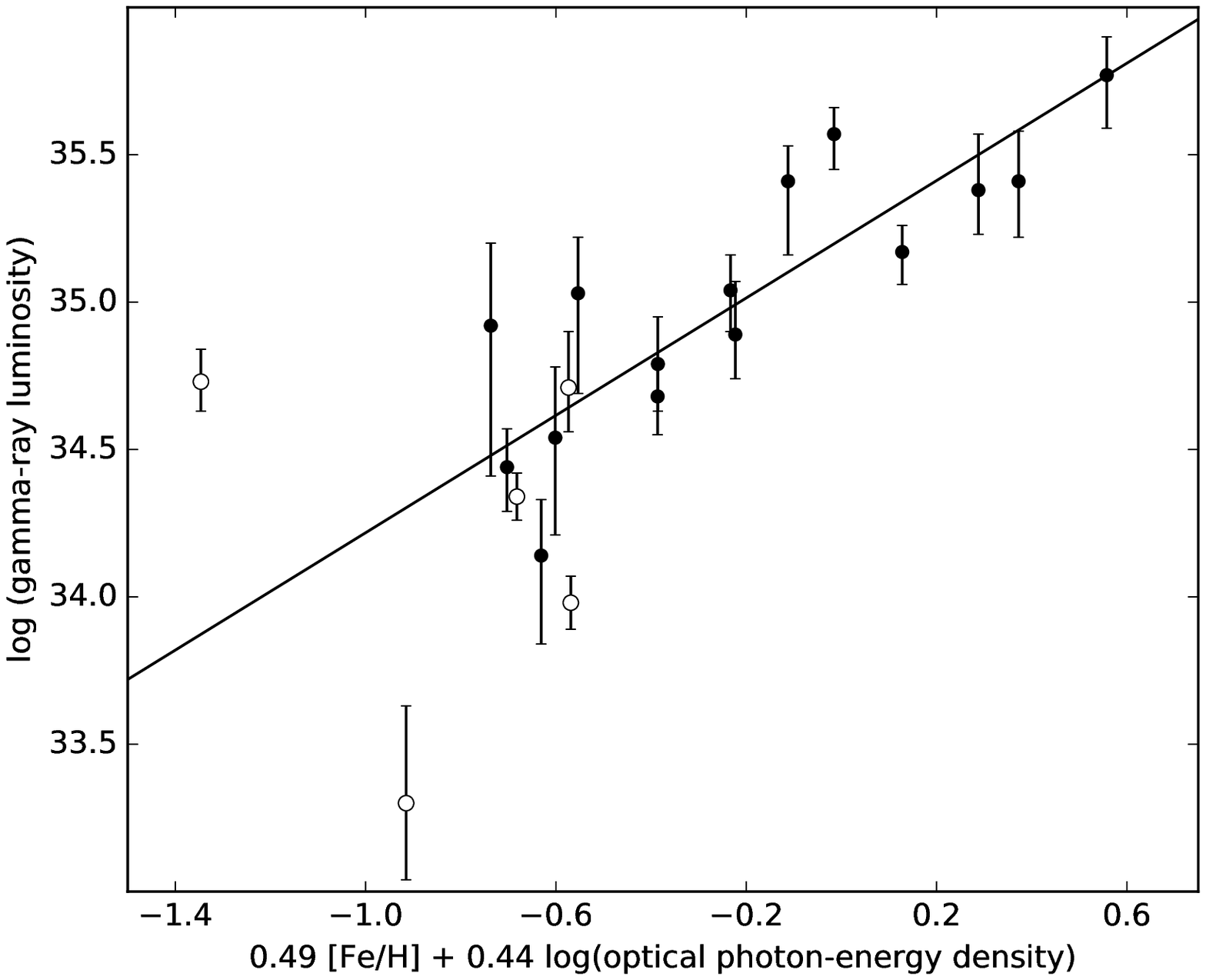}
		\caption{The fundamental-plane relationships of gamma-ray GCs. The filled circles are adopted from \citet{hui} and the open circles represent the five GCs studied in this work. The solid line represents the best-fit of the fundamental-plane relationships (adopted from \citet{hui}).}
	\label{plane}
\end{figure*}

The detection prospect of gamma-ray emission from GCs depends on their gamma-ray luminosities (i.e. the $\dot{E}$ and gamma-ray efficiency) as well as their locations. It is generally believed that $L_\gamma$ correlates with the encounter rate $\Gamma_{c}$, the metallicity [Fe/H] \citep{hui}, the number of MSPs, the shape of spectrum of source and the distance from us. In the inverse Compton scattering model of gamma-ray GCs, $L_\gamma$ is governed by the energy density of the soft-photon field at the cluster location (see \citet{cheng2}). To further check these possibilities, we have a plot (i.e., Figure~\ref{plane}) following \citet{hui}. The current five gamma-ray GCs fit in the fundamental-plane relationships suggested in \citet{hui}. An improved sample of $\gamma$-ray GCs can further test the correlation between the detected gamma-ray luminosities and other properties of GCs, which may shed additional light on the origin of GeV gamma-ray photons. For example, the tendency that the gamma-ray luminosities increase with the energy densities of the soft photons, as shown in Figure \ref{plane}, is in support of the IC scattering mechanism \citep{cheng2}. Moreover, the locations of the GCs are important. The higher the Galactic latitude, the lower the gamma-ray background. Hence the sources at high Galactic latitudes are more easily to identify. That is why the first group of gamma-ray GCs were identified at relatively high Galactic latitudes while the sources to be discovered were mainly in the Galactic plane. Nevertheless, the gamma-ray GC sample is expected to grow up due to the continual successful operation of Fermi-LAT and the ongoing observation of Dark Matter Particle Explorer \citep{Chang2014}. With a significantly enhanced sample, some fundamental differences between the gamma-ray GCs and non-detected ones, if there are, will be identified.

\section*{ACKNOWLEDGEMENTS}
We acknowledge the use of data from FSSC and SIMBAD database and thank the anonymous referee for the insightful comments/suggestions. We also thank Dr. Yizhong Fan for helpful discussion. This work was supported in part by 973 Program of China under grant 2013CB837000, the National Natural Science Foundation of China under grants 11273064, 11433009 and 11573071, and the Strategic Priority Research Program of CAS (under grant number XDB09000000).

\clearpage
\begin{table*}
\begin{center}
\caption{Parameters of gamma-ray globular clusters.}
\begin{tabular}{ccrrrccrcccc}
\hline\hline
GC Name& \multicolumn{4}{c}{Center of ROI$^{(a)}$} & Tidal radius$^{(b)}$ & \multicolumn{3}{c}{Position and error $^{(c)}$} & Offset$^{(d)}$& Distance & Energy band \\
                  & R.A. & Dec. & {\it l.} & {\it b.} & arcmin & R.A. & Dec. & radian & arcmin & kpc &(GeV)  \\
\hline
 NGC 5904 & 229.64 & 2.08    & 3.86    &  46.80 & 12.07  & 229.63 & 2.04 & $\pm$0.06 & 2.65 & 7.5$^{(e)}$ & 0.1-500\\
 M 15           & 322.35 & 12.1    & 65.01  &  27.31 &  21.50 & 322.44 & 12.13 & $\pm$0.15 & 3.72 & 10.3$^{+0.4(f)}_{-0.4}$ & 0.1-500   \\
 NGC 6397 & 265.18 &$-$53.67 & 338.17& $-$11.96 & 15.81 & 265.30 & $-$53.66 & $\pm$0.06 & 4.40 & 2.2$^{+0.5(g)}_{-0.7}$ & 0.2-500	\\
 NGC 6218 & 251.81 &$-$1.95   &15.72    & 26.31   & 17.6   & 251.82 & $-$1.89 & $\pm$0.04 & 3.68 & 4.8$^{(e)}$ & 0.1-500 \\
 NGC 6139 & 246.92 &$-$38.85 &342.37 & 6.94      & 8.52   & 246.92 & $-$38.88 & $\pm$0.05 & 1.73 & 9.4$^{+0.5(h)}_{-0.7}$ & 0.1-500\\
\hline
\end{tabular}
\label{para_source}
\end{center}
{\bf Notes.}
(a) Coordinates derived from \citet[][ 2010 version]{harris} in J2000 and their Galactic coordinates. (b) Tidal radius of five globular clusters. (c)The best-fit position and 1$\sigma$ error of results of {\it gtfindsrc}. (d) The offsets from their cores. (e) Distance from the Sun derived from \citet[][ 2010 version]{harris}. (f) Derived from \citet{van_den}. (g) Derived from \citet{heyl}. (h) Derived from \citet{ortolani}.
\end{table*}

\begin{table*}
\begin{center}
\caption{Results of gamma-ray globular clusters.}
\begin{tabular}{cccccccccccc}
\hline\hline
GC  & Spectral  & TS & Significance & Photon & Cutoff& photon$^{(1)}$  & Energy$^{(2)}$ & Luminosity$^{(3)}$ & N$_{MSP}^{(4)}$&  ${\cal N}_{MSP}^{(5)}$\\
Name & model &  & (post-trial;$\sigma$) & index & (GeV) & flux  & flux &    &    & &\\
\hline
\multicolumn{12}{c}{Five GCs for this paper}\\\hline
 NGC 5904 & PL  &  44  & 5.3 & 2.40 $\pm$ 0.14 &...& 5.94 $\pm$ 1.87  & 3.22 $\pm$ 0.60  & 2.18 $\pm$ 0.40    & 15 $\pm$ 3 & 5  \\
                     & PLE & 43 & 4.9 & 1.82 $\pm$ 0.50 & 4.0 $\pm$ 3.2 & 3.32 $\pm$ 2.14 & 2.24 $\pm$ 0.71 & 1.51 $\pm$ 0.48 & 11 $\pm$ 4 & 5 \\
 M 15           & PL  &  49  & 5.7 & 2.84 $\pm$ 0.18 &...& 11.84 $\pm$ 2.48 & 4.15 $\pm$ 0.65  & 5.26 $^{+~1.31}_{-~1.16}$   & 37 $^{+~9}_{-~8}$  & 8 \\
 NGC 6397 & PL &  49  & 5.7 & 2.83 $\pm$ 0.17 &...& 4.94 $\pm$ 0.88 & 3.47 $\pm$ 0.54 & 0.20 $^{+~0.15}_{-~0.12}$ & 2 $\pm$ 1  & 1 \\
 NGC 6218 & PL &  36 & 4.5 & 2.28 $\pm$ 0.20 &...& 5.28 $\pm$ 2.44 & 3.46 $\pm$ 0.73  & 0.95 $\pm$ 0.20 & 7 $\pm$ 2  & None \\
 NGC 6139 & PL &  27 & 3.4 & 2.44 $\pm$ 0.12 &...& 9.56 $\pm$ 2.42 & 4.93 $\pm$ 0.86  & 5.21 $^{+~2.28}_{-~1.78}$ & 36 $^{+~16}_{-~13}$ & None \\\hline
 \multicolumn{12}{c}{Previous results}\\\hline
 NGC 5904$^{(6)}$ & PL  &  26  & 3.2 & 2.3 $\pm$ 0.2 &...& 6.2 $\pm$ 1.4  & 3.5 $\pm$ 0.8  & 2.4 $\pm$ 0.5    & 16 $\pm$ 4 & ...  \\
 M 15$^{(7)}$ & PLE  &  5.4  & ... & (1.4) & (1.6) & $<$6.0  & $<$5.0  & $<$5.8    & $<$56 & ... \\
 NGC 6139$^{(8)}$ & PL  &  32  & ... & 2.1 $\pm$ 0.2 &...& 9.9 $\pm$ 5.4  & 8.8 $\pm$ 4.8  & 11 $\pm$ 6    & 75 $^{+114}_{-50}$ & ...  \\\hline
\end{tabular}
\label{para_spec}
\end{center}
{\bf Notes.}
(1) Integrated photon flux in unit of 10$^{-9}$ cm$^{-2}$ s$^{-1}$. (2) Integrated energy flux in unit of 10$^{-12}$ erg cm$^{-2}$ s$^{-1}$. (3) Luminosity in unit of 10$^{34}$~erg~s$^{-1}$. (4) The evaluated number of MSPs N$_{MSP}$. (5) The detected number of MSPs ${\cal N}_{MSP}$. (6) The data adopted from \citet{zhou}. (7) The data adopted from \citet{abdo2010c}. (8) The data adopted from \citet{tam}.
\end{table*}
\clearpage
\begin{table*}
\begin{center}
\caption{Properties of Gamma-ray globular clusters.}
\begin{tabular}{lccccccc}
\hline\hline
Cluster Name & $d^{(1)}$ & $\Gamma_{\rm c}^{(2)}$ & [Fe/H]$^{(3)}$ & $M_V^{(4)}$ & $u_{\rm optical}^{(5)}$ & $u_{\rm IR}^{(5)}$ & $\log L_{\gamma}^{(6)}$ \\
{}  & (kpc) &  &  & & (eV~cm$^{-3})$ & (eV~cm$^{-3})$ & (erg~s$^{-1})$ \\\hline\hline
\multicolumn{8}{c}{Fifteen GCs for \citet{abdo2010c} and \citet{tam}}\\\hline
47 Tuc$^{(7)}$  & 4.0 & 44.13 & -0.76 & -9.17 & 0.93 & 0.25 & $34.68^{+0.12}_{-0.13}$\\
Omega Cen$^{(7)}$ & 4.8 & 4.03 & -1.62 & -10.07 & 1.61 & 0.51 & $34.44^{+0.13}_{-0.15}$\\
M 62$^{(7)}$  & 6.6 & 47.15 & -1.29 & -9.09 & 8.07 & 0.86 &  $35.04^{+0.12}_{-0.14}$\\
NGC 6388$^{(7)}$  & 11.6 & 101.99 & -0.60 & -9.74 & 2.59 & 0.56 & $35.41^{+0.12}_{-0.25}$\\
Terzan 5$^{(7)}$  & 5.5 & 118.29 & 0.00 & -6.51 & 7.02 & 1.37 & $35.41^{+0.17}_{-0.19}$\\
NGC 6440$^{(7)}$  & 8.5  & 74.17 & -0.34 & -8.78 & 10.79 & 1.00 & $35.38^{+0.19}_{-0.15}$\\
M 28$^{(7)}$  & 5.1 & 13.10 & -1.45 & -7.98 & 5.47 & 0.92 & $34.79^{+0.16}_{-0.16}$\\
NGC 6652$^{(7)}$  & 9.0 & 1.24 & -0.96 & -6.43 & 3.65 & 0.51 & $34.89^{+0.18}_{-0.15}$\\\hline
Liller 1$^{(8)}$  & 9.6 & 77.98  & 0.22 & -7.63 & 10.53 & 1.40 & $35.77^{+0.13}_{-0.18}$  \\
M 80$^{(8)}$      & 10.3 & 31.31 & -1.75 & -8.23 & 1.88 & 0.33 & $34.92^{+0.28}_{-0.51}$  \\
NGC 6441$^{(8)}$  & 11.7 & 88.42 & -0.53 & -9.64 & 3.59 & 0.69 & $35.57^{+0.09}_{-0.12}$  \\
NGC 6624$^{(8)}$  & 7.9 & 14.65 & -0.44 & -7.49 & 6.03 & 0.67 &  $35.17^{+0.09}_{-0.11}$ \\
NGC 6541$^{(8)}$  & 6.9 & 20.00 & -1.83 & -8.34 & 4.69 & 0.61 &  $34.54^{+0.24}_{-0.33}$ \\
NGC 6752$^{(8)}$  & 4.4 & 10.78 & -1.56 & -7.94 & 2.01 & 0.48 &  $34.14^{+0.19}_{-0.30}$ \\
NGC 6139$^{(8)}$  & 10.1 & 13.28 & -1.68 & -8.36 & 4.10 & 0.69 & $35.03^{+0.19}_{-0.34}$  \\\hline
\multicolumn{8}{c}{Five GCs for this paper}\\\hline
NGC 5904  & 7.5 & 5.11 & $-$1.29 &-8.81& 0.77 &  0.17 & 34.34 $\pm$ 0.08\\
M 15            &10.4&11.85& $-$2.37 &-9.19& 0.38 &  0.10 & 34.73 $^{+~0.11}_{-~0.10}$\\
NGC 6397  & 2.3 & 1.63 & $-$2.02 &-6.64& 1.48 &  0.57 & 33.30 $^{+~0.33}_{-~0.26}$\\
NGC 6218  & 4.8 & 0.56 & $-$1.37 &-7.31& 1.71 &  0.41 & 33.98 $^{+~0.09}_{-~0.09}$\\
NGC 6139  &10.1&14.07& $-$1.65 &-8.36&3.42 &  0.65 & 34.71 $^{+~0.19}_{-~0.15}$\\
\hline\hline
\end{tabular}
\label{proper_source}
\end{center}
{\bf Notes.}
(1) Distance from the Sun derived from \citet[][ 2010 version]{harris}.
(2) The parameter (encounter rate) was estimated by $\rho_{0}^{2}r_{c}^{3}\sigma_{0}^{-1}$ \citep[cf.][]{hui}. The value of $\rho_{0}$, $\sigma_{0}$ and r$_{\rm c}$ derived from \citet[][ 2010 version]{harris} and \citet{gnedin}.
(3) Metallicity of each GC.
(4) Absolute visual magnitude derived from \citet[][ 2010 version]{harris}.
(5) Energy densities of various soft photon fields \citep[cf.][]{strong}.
(6) $\gamma-$ray luminosities of GCs.
(7) The luminosities adopted from \citet{abdo2010c}.
(8) The luminosities adopted from \citet{tam}.
\end{table*}
\clearpage
\begin{figure*}
\centering
	\includegraphics[scale=0.4]{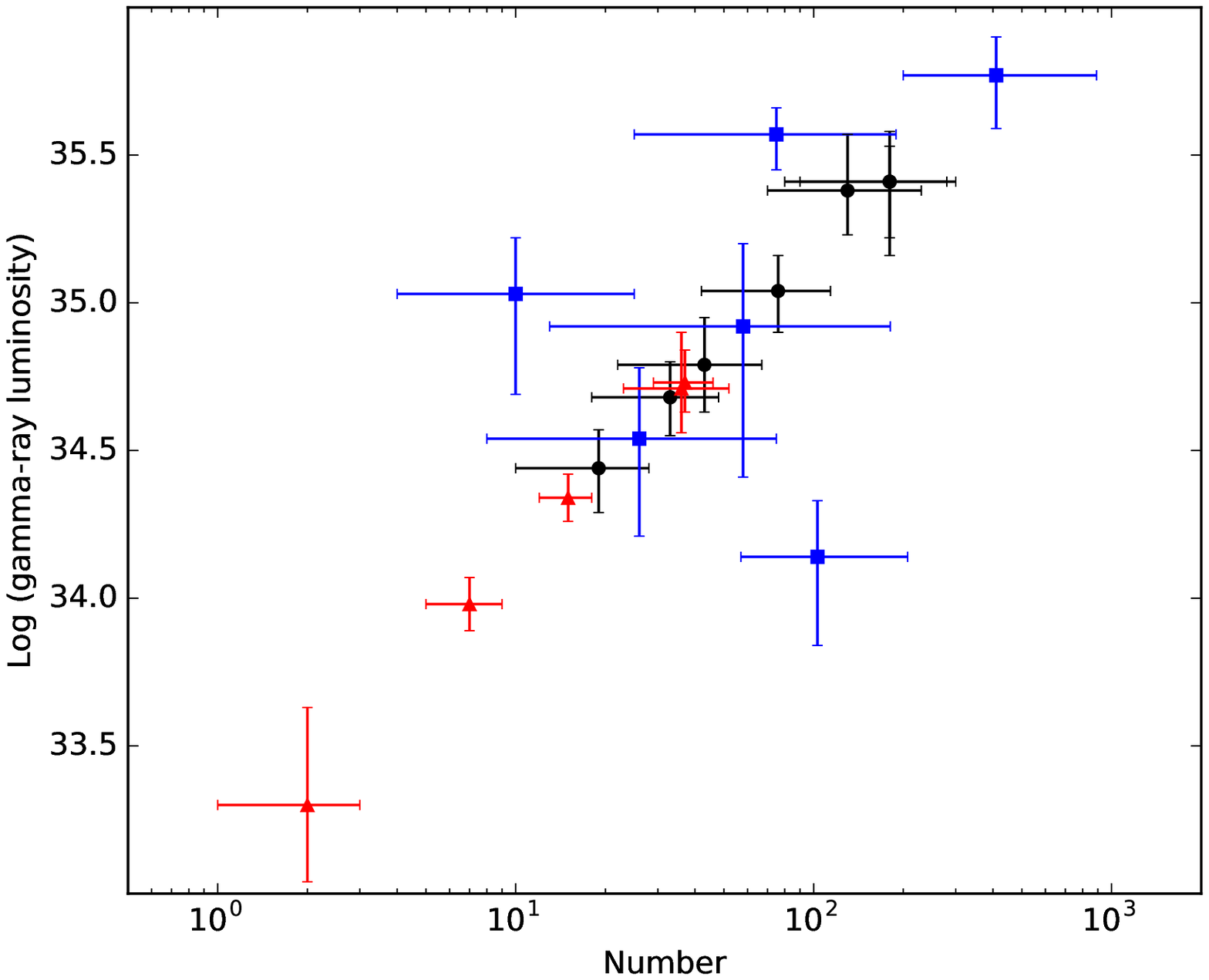}
	\includegraphics[scale=0.4]{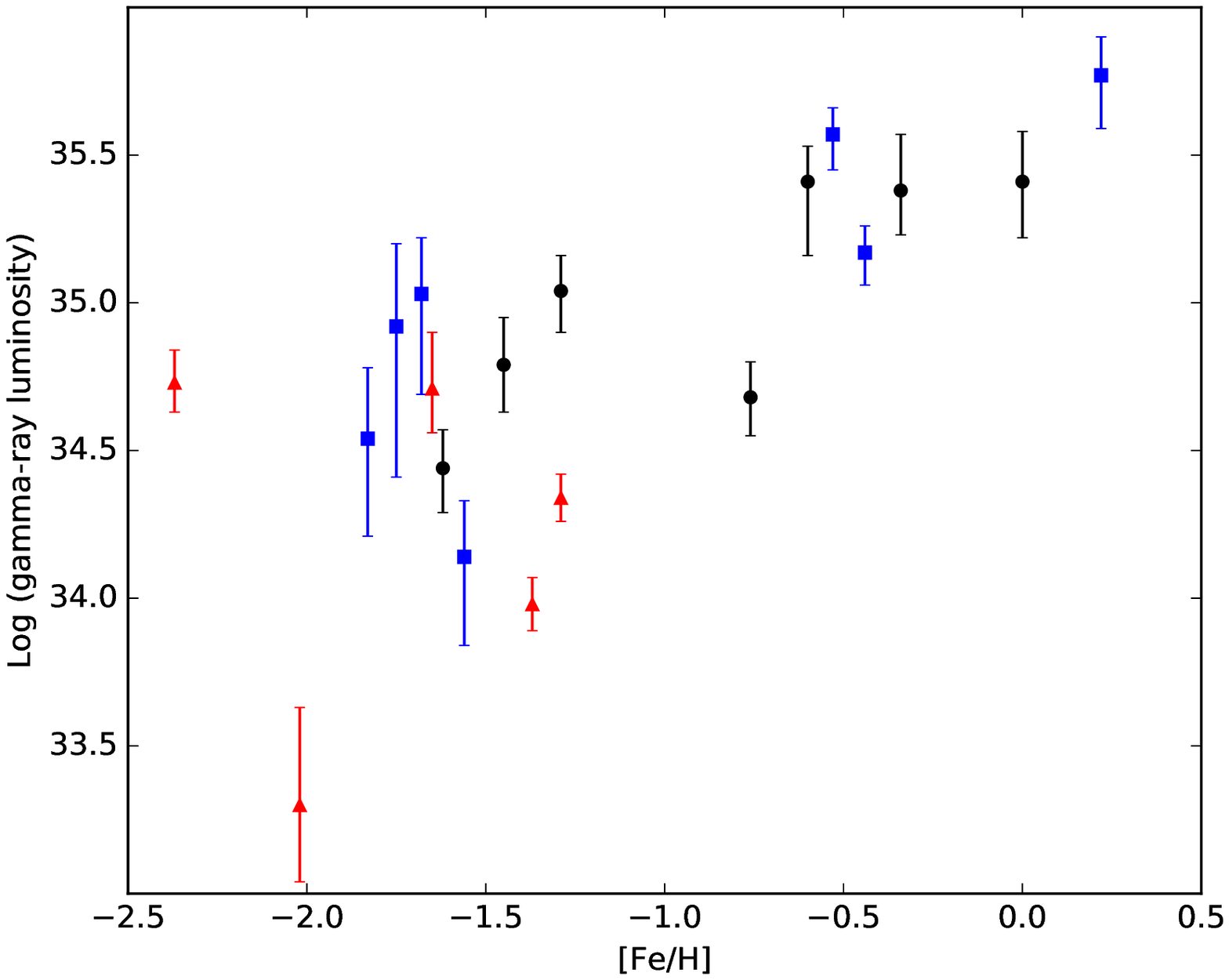}
	\includegraphics[scale=0.4]{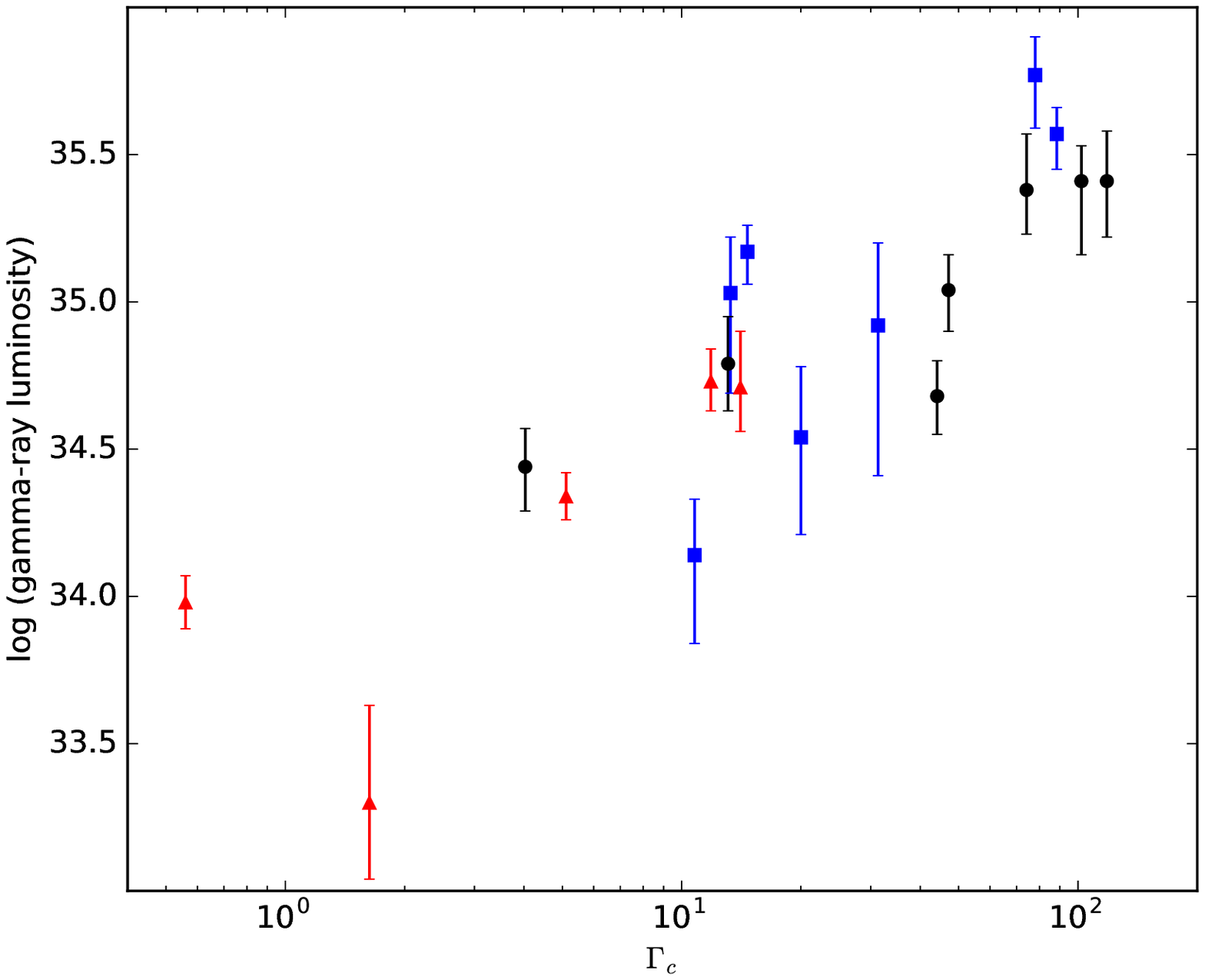}
	\includegraphics[scale=0.4]{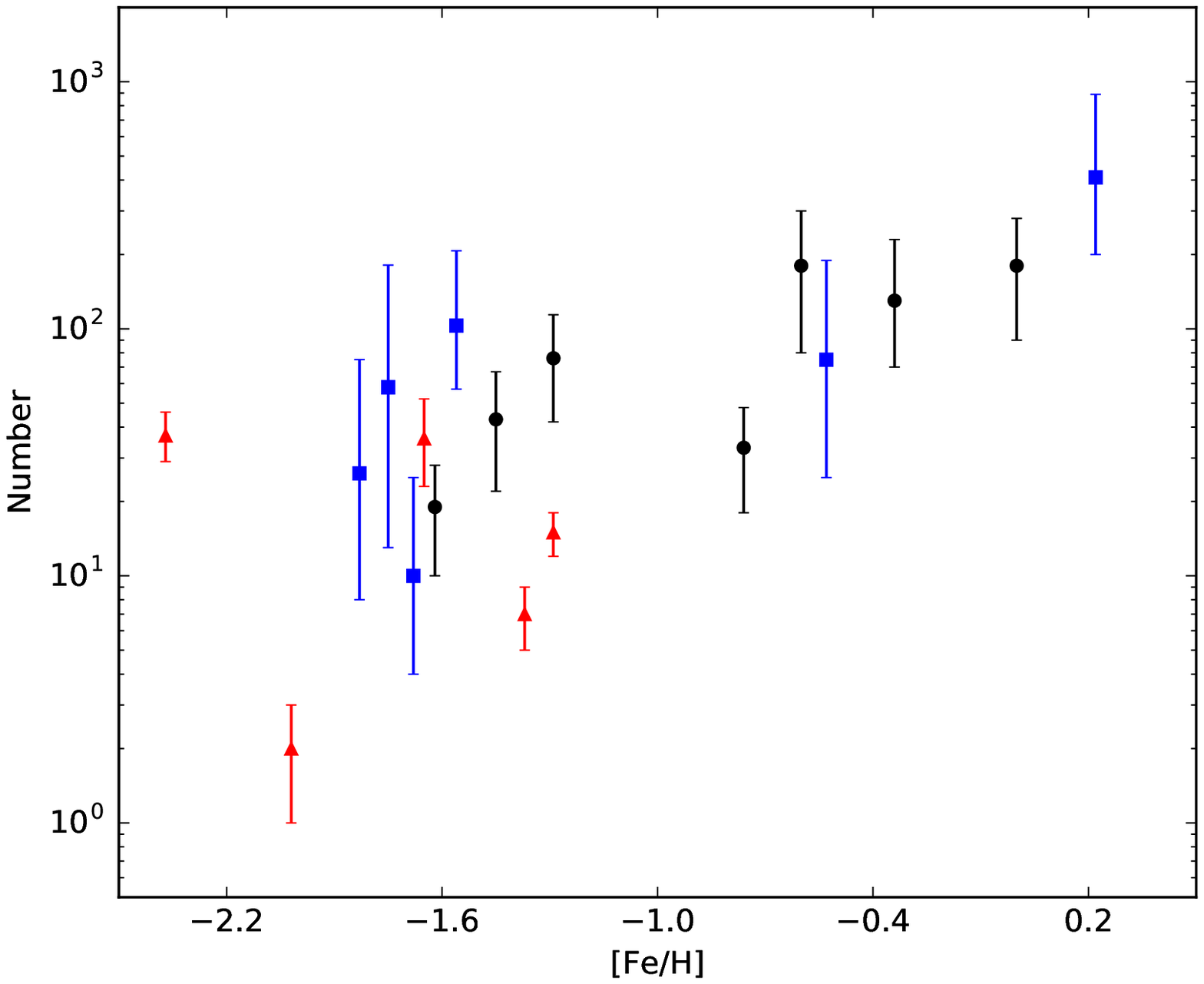}
	\includegraphics[scale=0.4]{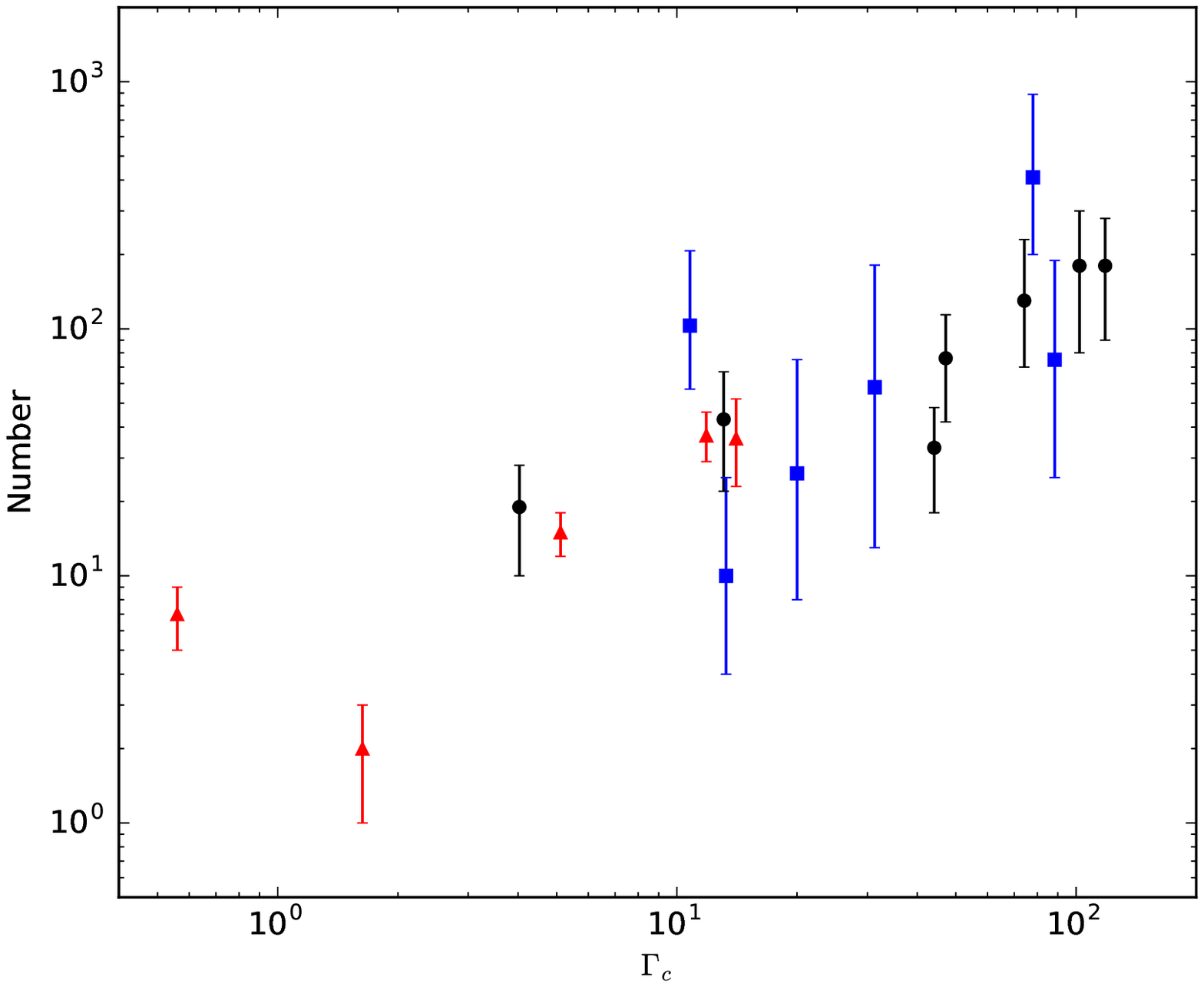}
	\includegraphics[scale=0.4]{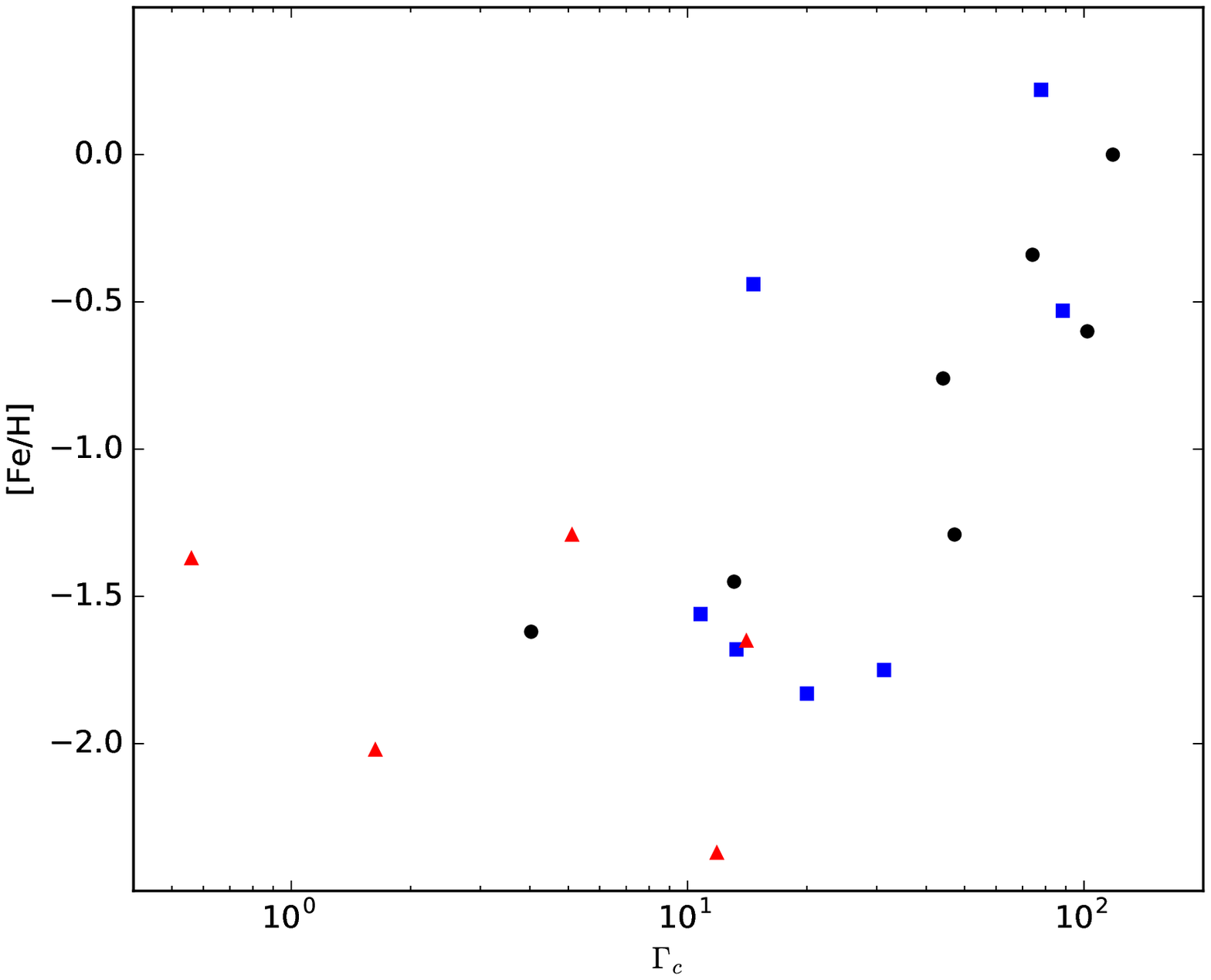}

		\caption{Comparisons of gamma-ray luminosities, the number of MSPs, encounter rate ($\Gamma_{c}$) and metallicity [Fe/H] between the previous results and ours. The black points, blue squares and red triangles represent the data adopted from \citet{abdo2010c}, \citet{tam} and our own, respectively.}
	\label{par_vs}
\end{figure*}
\label{lastpage}

\begin{thebibliography}{}
\bibitem[Abdo et al. (2009a)]{abdo2009a}Abdo, A. A., et al., 2009a, Science, 325, 848
\bibitem[Abdo et al. (2009b)]{abdo2009b}Abdo, A. A., et al., 2009b, Science, 325, 845
\bibitem[Abdo et al. (2010a)]{abdo2010a}Abdo, A. A., et al., 2010a, Science, 328, 725
\bibitem[Abdo et al. (2010b)]{abdo2010b}Abdo, A. A., et al., 2010b, ApJS, 188, 405
\bibitem[Abdo et al. (2010c)]{abdo2010c}Abdo, A. A., et al., 2010c, A\&A, 524, 75
\bibitem[Acero et al. (2015)]{acero}Acero, F., et al. 2015, ApJS, 218, 23
\bibitem[Anderson et al. (1997)]{anderson}Anderson, S.~B., Wolszczan, A., Kulkarni, S.~R., Prince, T.~A., 1997, ApJ, 482, 870
\bibitem[Atwood et al. (2009)]{Atwood2009}Atwood, W. B., et al. 2009,  ApJ, 697, 1071
\bibitem[Alpar et al. (1982)]{alpar}Alpar, M. A., Cheng, A. F., Ruderman, M. A., Shaham, J., 1982, Nature, 300, 728
\bibitem[Bednarek \& Sitarek (2007)]{bednarek}Bednarek, W. \& Sitarek, J., 2007, MNRAS, 377, 920
\bibitem[Chang (2014)]{Chang2014} Chang, J. 2014, Chin. J. Spac. Sci., 34, 550 (http://www.cjss.ac.cn/CN/10.11728/cjss2014.05.550)
\bibitem[Cheng \& Taam (2003)]{cheng}Cheng, K., S., \& Taam, R., E., 2003, ApJ, 300, 500
\bibitem[Cheng et al. (2010)]{cheng2}Cheng, K. S., Chernyshov, D. O., Dogiel, V. A., Hui, C. Y., and Kong, A. K. H., 2010, ApJ, 723, 1219
\bibitem[Clark (1975)]{clark}Clark, G. W., 1975, ApJL, 199, L143
\bibitem[Edwards et al. (2006)]{edwards}Edwards, R. T., Hobbs, G. B., \& Manchester, R. N. 2006, MNRAS, 372, 1549
\bibitem[Freire et al. (2011)]{freire}Freire, et al., 2011, Science, 334, 1107
\bibitem[Gnedin et al. (2002)]{gnedin}Gnedin, O.~Y., Zhao, H.~S., Pringle, J. E., Fall, S. M., Livio, M., and Meylan, G., 2002, ApJ, 568, L23
\bibitem[Gratton et al. (2003)]{gratton}Gratton, R. G., Bragaglia, A., Carretta, E., Clementini, G., Desidera, S., Grundahl, F., Lucatello, S., 2003, A\&A, 408, 529
\bibitem[Harding et al. (2005)]{harding}Harding, Alice K., Usov, Vladimir V., Muslimov, Alex G., 2005, ApJ, 622, 531
\bibitem[Harris (1996)]{harris}Harris, W.E. 1996, AJ, 112, 1487
\bibitem[Hessels et al. (2007)]{hessels}Hessels, J.~W.~T., Ransom, S.~M., Stairs, I.~H., Kaspi, V.~M., Freire, P.~C.~C., 2007, ApJ, 670, 363
\bibitem[Heyl et al. (2012)]{heyl}Heyl, J. S., Richer, H., Anderson, J., Fahlman, G., Dotter, A., Hurley, J., Kalirai, J., Rich, R. M., Shara, M. \& Stetson, P., 2012, ApJ, 761, 51
\bibitem[Hui et al. (2011)]{hui}Hui, C.~Y., Cheng, K.~S., Wang, Y., Tam, P. H. T., Kong, A. K. H., Chernyshov, D. O., and Dogiel, V. A., 2011, ApJ, 726, 100
\bibitem[Hobbs et al. (2006)]{hobbs}Hobbs, G. B., Edwards, R. T., \& Manchester, R. N., 2006, MNRAS, 369, 655
\bibitem[Katz (1975)]{katz}Katz, J. I., 1975, Nature, 253, 698
\bibitem[Kong et at. (2010)]{kong}Kong, A. K. H., Hui, C. Y., Cheng, K. S., 2010, ApJL, 712, 36
\bibitem[Nolan et al. (2012)]{nolan2012}Nolan, P. L., et al. 2012, ApJS, 199, 31
\bibitem[Ortolani et al. (1999)]{ortolani}Ortolani, S., Bica2, E., and Barbuy, B., 1999, A\&A, 138, 267
\bibitem[Strong \& Moskalenko. (1998)]{strong}Strong, A. W., \& Moskalenko, I. V., 1998, ApJ, 509, 212
\bibitem[Tam et al. (2011)]{tam}Tam, P. H. T., Kong, A. K. H., Hui, C. Y., Cheng, K. S., Li, C., \& Lu, T.-N., 2011, ApJ, 729, 90
\bibitem[van den Bosch et al. (2006)]{van_den}van den Bosch, R. C. E., de Zeeuw, P. T., Gebhardt, K., Noyola, E., \& van de Ven, G. 2006, ApJ, 641, 852
\bibitem[Venter et al. (2008)]{venter}Venter, C., de Jager, O. C., 2008, ApJ, 680, L125
\bibitem[Verbunt \& Hut (1987)]{verbunt}Verbunt, F., Hut, P., 1987, IAU Symp. 125, 187
\bibitem[Wei et al. (1996)]{wei}Wei, D. M., Cheng, K. S., Lu, T., 1996, ApJ, 468, 207
\bibitem[Zhang et al. (2003)]{zhang}Zhang, L. \& Cheng, K. S., 2003, A\&A, 398, 639
\bibitem[Zhou et al. (2015)]{zhou}Zhou, J.~N., Zhang, P.~F., Huang, X.~Y., Li, X., Liang, Y.~F., Fu, L., Yan, J.~Z., \& Liu, Q.~Z., 2015, MNRAS, 448, 3215

\end{thebibliography}
\end{document}